\documentclass[12pt,letterpaper]{article}
\pdfoutput=1

\usepackage{graphicx}
\usepackage{geometry}
\usepackage{amsmath,amsthm,amsfonts,amssymb,amscd}
\usepackage{xcolor}
\usepackage{setspace}
\usepackage{cite}
\usepackage{cancel}
\usepackage{cases}

\definecolor{blue2}{rgb}{0.15,0.15,0.8}

\usepackage[linktocpage=true]{hyperref}
\hypersetup{
colorlinks=true,
citecolor=blue2,
linkcolor=blue2,
urlcolor=blue2,
pdfauthor={},
pdftitle={},
pdfsubject={}
}

\newcommand{\GSM}{{\rm G}_{\rm SM}}
\newcommand{\SU}{\,{\rm SU}}
\newcommand{\U}{\,{\rm U}}
\newcommand{\UoneD}{{\rm U}(1)_{\rm D}}
\newcommand{\UonePQ}{{\rm U}(1)_{\rm PQ}}

\begin{document}

\thispagestyle{empty}

\begin{center}
\begin{doublespace}
{\LARGE\bf Integer solutions to the anomaly equations \\for a class of chiral gauge theories}\\
\end{doublespace}
\vspace{0.8cm}
{\Large Alessandro Podo$^a$, Filippo Revello$^b$} 
\end{center}

\begin{center}
{$^a$\small \textit{Department of Physics, Center for Theoretical Physics, \\Columbia University, New York, 538 West 120th Street, NY 10027, USA}\\
\vspace{0.3cm}
$^b$\small \textit{Rudolf Peierls Centre for Theoretical Physics, Beecroft Building, Clarendon Laboratory,\\ Parks Road, University of Oxford, OX1 3PU, UK}\\}
\end{center}

\begin{abstract} \noindent 
We find all the integer charge solutions to the equations for the cancellation of local gauge anomalies in a class of gauge theories which extend the Standard Model (SM) by a gauge group of the form $G \times \U(1)$, where $G$ is an arbitrary semisimple compact Lie group. The SM fermions are assumed to be neutral under $G \times \U(1)$ gauge interactions, while the new fermions transform in nontrivial representations of both the new and the SM gauge groups. Our analysis is valid also when the latter is embedded in an arbitrary semisimple compact Lie group. Theories with this structure have been recently studied as models of composite axions based on accidental symmetries and can provide a field theory resolution to the axion quality problem. We apply our results to cases of phenomenological interest and prove the existence of charge assignments with Peccei--Quinn symmetry protected up to dimension 18.
\end{abstract}

\newpage

\section{Introduction}
The consistency of quantum gauge theories relies on the cancellation of gauge anomalies, see for instance \cite{Weinberg:1996kr}. These conditions impose nontrivial restrictions on the choice of charges and representations for fermions with vanishing mass. This is particularly relevant for chiral gauge theories, in which the fermionic mass terms are forbidden by gauge invariance. In the case of four dimensional theories and in the presence of a $\U(1)$ gauge group, the condition for the vanishing of the $[\U(1)]^3$ anomaly, corresponding to a triangle diagram, translates into an homogeneous cubic equation for the charges of the form: 
$$
x_1^3 + \dots + x_n^3=0.
$$
Additional conditions derive from the cancellation of mixed gauge anomalies.

If the charges are integers\footnote{Of course up to an overall redefinition of the $\U(1)$ gauge coupling constant.}, as believed to be the case if the theory admits a quantum gravity UV completion~\cite{Banks:2010zn,Harlow:2018tng} and mandatory if the $\U(1)$ group is embedded in a simple non-Abelian gauge group, then one has to deal with a system of Diophantine equations, \emph{i.e.} polynomial equations over the ring of integers. Cubic Diophantine equations are an active research topic in mathematics and no general method of solution is known at present times. Finding the integer solutions to the anomaly equations thus appears a formidable task. 

In the past few years, however, starting from the seminal work of Ref.~\cite{Costa:2019zzy} (see also \cite{Allanach:2018vjg} for earlier progress), it has been realised that some explicit cases of physical relevance can be solved completely using elementary methods. The solved cases include purely Abelian gauge theories~\cite{Costa:2019zzy,Allanach:2019gwp,Costa:2020dph}, Abelian extensions of the Standard Model (SM)~\cite{Allanach:2019uuu,Allanach:2020zna,Costa:2020krs,Dobrescu:2020evn} and even the gauge-only\footnote{That is, without imposing the cancellation of mixed gauge-gravitational anomalies.} anomalies of the SM \cite{Lohitsiri:2019fuu}. To the best of our knowledge no result is currently available for more general non-Abelian extensions of the SM with fixed matter content. Scenarios based on these constructions have been extensively studied with several phenomenological motivations, including composite Higgs models~\cite{Kaplan:1983fs,Kaplan:1983sm,Dugan:1984hq} (see for instance~\cite{Contino:2010rs,Panico:2015jxa} for more recent reviews), models of composite QCD axion~\cite{Kim:1984pt,Kaplan:1985dv,Choi:1985cb,Randall:1992ut,Dobrescu:1996jp,Redi:2016esr,Fukuda:2017ylt,Lillard:2017cwx,Lillard:2018fdt,Gavela:2018paw,Vecchi:2021shj,Contino:2021ayn}, and studies of dark matter and dark sectors~\cite{Strassler:2006im,Han:2007ae,Arkani-Hamed:2008hhe,Kilic:2009mi,Falkowski:2009yz,Bai:2010qg,Buckley:2012ky,Boddy:2014yra,Antipin:2015xia,Harigaya:2016rwr,Co:2016akw,Mitridate:2017oky,Contino:2018crt,Contino:2020god} (see~\cite{Kribs:2016cew,Cline:2021itd} for recent reviews). Solving the corresponding anomaly equations provides a way to systematically classify them and restrict the space of physically admissible theories. In this work we shall take one step in this direction by providing the general solution to the equations for the cancellation of local gauge anomalies in a class of chiral gauge theories that extend the SM with a non-Abelian gauge group of the form $G \times \U(1)$, where $G$ is a semisimple compact Lie group. Our methods of solution will be along the lines of the enlightening geometric approach described in Ref.~\cite{Allanach:2019gwp}.

We consider theories that extend the SM matter content by a fixed number of dark fermions with charge assignments that are vectorlike under $G \times G_{\rm SM}$, but otherwise arbitrary under the $\U(1)$ factor. The analysis is valid also when the SM factor is embedded in an arbitrary semisimple compact Lie group, making it compatible with all the most popular Grand Unification scenarios. Hidden valley models~\cite{Strassler:2006im,Han:2007ae}, where the SM is augmented by a non-Abelian confining $\SU(N)$ factor, offer a compelling mechanism to decouple visible physics from a dark sector at relatively low energy scales. In the hidden valley scenario, the dark fermions are not charged under the SM, but the existence of an Abelian $\U(1)$ factor in addition to the confining factor $\SU(N)$ enables a portal between the two sectors, through the kinetic mixing of the $\U(1)$ and the SM hypercharge gauge bosons. On the other hand, it is also possible to consider the case where the fundamental dark constituents are in fact charged under the SM, but the confining dynamics of $G$ leaves the SM unbroken. This paradigm, known as vectorlike confinement \cite{Kilic:2009mi}, has received considerable interest in the context of beyond the SM phenomenology. Composite states of the dark gauge theory --- such as baryons, pions, or more exotic states --- can be stable thanks to accidental global symmetries and act as promising dark matter candidates \cite{Falkowski:2009yz,Bai:2010qg,Buckley:2012ky,Boddy:2014yra,Antipin:2015xia,Harigaya:2016rwr,Co:2016akw,Mitridate:2017oky,Contino:2018crt,Contino:2020god}. The coupling of the dark fermions to $G_{\rm SM}$, moreover, is a necessary ingredient in the models of composite QCD axions. One might wonder if a dark $\U(1)$ could still play a role in these cases, even if the two sectors are already in communication through SM gauge interactions. An interesting possibility is that the additional $\U(1)$ makes the matter content chiral, despite a vectorlike assignment under $G \times G_{\rm SM}$. If the $\U(1)$ gauge coupling is perturbative, this allows one to have a chiral gauge theory while still keeping control of the low energy dynamics of the non-Abelian factor $G$ (such as its confining nature and the pattern of chiral symmetry breaking). Gauge invariance then forbids mass terms for the fermions and the low energy phenomenology is completely determined by the dark confinement scale. Scenarios of this kind have been studied in\cite{Contino:2020god,Contino:2021ayn}, and give rise to a rich interplay between cosmological and laboratory probes.  In combination with theoretical constraints, for instance anomaly cancellation and perturbativity of the gauge couplings, such models can be classified with a minimal set of assumptions. In this spirit, we hope our results may be useful beyond the specific examples we consider, either directly or through an application of the same techniques to similar problems.

As a nontrivial application of our results, we shall consider the theories introduced in Ref.~\cite{Contino:2021ayn} as models of composite axions with accidental Peccei--Quinn (PQ) symmetry, whose properties and phenomenological interest will be reviewed in Section~\ref{sec:composite_axions}. They can provide a resolution to the QCD axion quality problem by forbidding local PQ violating operators up to very high dimensions, depending on the integer values of the charges. They generalise previous constructions of models beyond the SM with accidentally stable composite dark matter candidates~\cite{Harigaya:2016rwr,Co:2016akw,Contino:2020god}. Our general solution for the anomaly equations can be useful, among other things, to assess which charge assignments give rise to a high quality QCD axion in the models of Ref.~\cite{Contino:2021ayn}.

The article is organised as follows.
In Section~\ref{sec:2} we present the class of gauge theories of interest, defining the matter content and gauge group structure, and formulate the mathematical problem for the cancellation of gauge anomalies. In Section~\ref{sec:3} we find all the integer solutions to the system of anomaly equations, building up from simple cases and then addressing the general case. The inclusion of SM hypercharges is then discussed in Section~\ref{sec:hypercharge}, before applying our findings to models of composite QCD axion that address the axion quality problem in Section~\ref{sec:composite_axions}. 
The reader interested only in the general solution to the anomaly equations can read the general formulation in Section~\ref{sec:2} and then jump directly to Section~\ref{sec:gen_sol}. 

\vspace{0.8cm}
\noindent
{\bf Notation and conventions.\\} 
Throughout the paper we work in $3+1$ dimensions. Given a unitary representation $r$ of a compact Lie group, we denote as $\bar{r}$ the complex conjugate representation.\\
We employ the notation $[x_1: \dots : x_n]$, with elements $x_i \in \mathbb{K}$ (the field $\mathbb{K}$ being $\mathbb{Q}$ or $\mathbb{R}$), to denote the equivalence classes of elements of $\mathbb{K}^n\backslash \{0\}$ under the equivalence relation: $(x_1, \dots , x_n)\sim (y_1, \dots , y_n)$ if there is a $\lambda \in \mathbb{K}$ such that $(x_1, \dots , x_n) = \lambda (y_1, \dots , y_n)$. This construction defines homogeneous coordinates on the projective space $\mathbb{PK}^{n-1}$.

\section{The models and the anomaly cancellation conditions}\label{sec:2}

We consider gauge theories which extend the SM by a gauge group of the form $G \times \U(1)$, where $G$ is a generic semisimple compact Lie group. For the sake of presentation we shall refer to gauge groups, but in fact our analysis will depend on the Lie algebra alone and we shall not specify the global gauge group structure.\\
The fermionic matter content of the theory is that of the SM, with three generations of quarks and leptons and the possible addition of right-handed neutrinos, which are all assumed to transform in the trivial representation of $G \times \U(1)$, plus a new set of fermions charged under $G \times \U(1) \times G_{\rm SM}$. In addition we assume that the new fermions have vectorlike charge assignments under both the non Abelian factor $G$ and the SM factor $G_{\rm SM}$. These requirements will be motivated more precisely in the rest of the section, and they correspond to the class of phenomenologically interesting models described in the introduction.

More in detail, we denote with $G_{\rm SM}$ the SM gauge group, or possibly a compact semisimple gauge group that extends it. This allows us to treat in full generality also the case in which the SM is embedded into a Grand Unification group at high energies (in $3+1$ dimensions). The case of a simple group corresponds to well-known Grand Unified Theories such as $\SU(5)$ or ${\rm Spin}(10)$, while more general semisimple extensions of the SM include also the Pati-Salam model $[\SU(4)\times \SU(2) \times \SU(2)]_{\rm PS}$ and have recently been classified in Ref.~\cite{Allanach:2021bfe}. 
We first consider the case in which $G_{\rm SM}$ is a generic semisimple gauge group which extends the SM, so that the equations for the cancellation of gauge anomalies simplify. The same analysis is valid also for the case in which $G_{\rm SM}$ is the SM gauge group $\SU(3)_{\rm QCD}\times \SU(2)_{\rm EW}\times \U(1)_Y$ but the new fermions carry only $\SU(2)_{\rm EW}$ and $\SU(3)_{\rm QCD}$ quantum numbers (\emph{i.e.} have zero hypercharge). The inclusion of hypercharge for the new fermions requires some care since it gives rise to additional anomaly conditions, and its analysis will be deferred to Section~\ref{sec:hypercharge}.

The models are defined in terms of two sets of left-handed Weyl (two-component) fermions $\{\psi_i\}$ and $\{\chi_i\}$, transforming under $G \times \U(1) \times G_{\rm SM}$ in representations:
\begin{equation}
\label{eq:genmod}
\psi_i \sim (R,p_i,r_i ), \qquad \chi_i \sim (\bar{R},q_i, \bar{r}_i), \quad \quad i = 1,...,n_{f},
\end{equation}
where $R$ and $r_i$ are generic, possibly reducible, finite dimensional unitary representations of $G$ and $G_{\rm SM}$ respectively, while $p_i$ and $q_i$ are integers labeling the $U(1)$ charges. We shall pay special attention to the cases $n_f=2$ and $n_f=3$, and provide a general solution for arbitrary $n_f$. 

It is convenient to introduce the notation:
\begin{equation}
\label{eq:dimT}
d_i = \dim(r_i) = \sum_\alpha \dim(r_{i,\alpha}), \qquad T_i = \sum_\alpha T(r_{i,\alpha}),
\end{equation}
where $r_i = \sum_\alpha r_{i,\alpha}$, with $r_{i,\alpha}$ irreducible, $\dim(r_{i,\alpha})$ is the dimension of each irreducible fragment and $T(r_{i,\alpha})$ is its Dynkin index.\footnote{We adopt a normalisation in which the Dynkin index is equal to $1$ for the fundamental representation, so that $T_i$ is always an integer.} We define similarly $d^{(G)}$ and $T^{(G)}$ as the dimension and the (total) Dynkin index of the representations $R,\bar{R}$ of $G$.

This class of models would correspond to a vectorlike set of fermions if the charges $p_i, q_i$ were vanishing. However, with a general charge assignment the model is chiral. From a phenomenological point of view, chiral models are particularly interesting, as mass terms for the dark fermions are forbidden by gauge invariance. All relevant scales in the IR (such as masses of the physical particles) are thus generated dynamically. This is in contrast to vectorlike models, where the fundamental fermions can have arbitrary mass terms. Despite being chiral, the theories we consider can give rise to an infrared dynamics that is under theoretical control, if the gauge group $G$ has a well-understood confining dynamics in the infrared, and if the $\U(1)$ and $G_{\rm SM}$ factors are weakly coupled at the confinement scale of $G$. The vectorlike structure with respect to $G_{\rm SM}$ ensures the existence of a $G_{\rm SM}$ preserving vacuum that is dynamically preferred in most cases~\cite{toappear}. A prominent example is obtained if $G=\SU(N)$ and the representation $R$ is the fundamental representation of $\SU(N)$. In many cases, depending on the fermion multiplicities, the infrared spectrum consists of a collection of (pseudo) Nambu--Goldstone bosons associated with the spontaneously broken global symmetries, plus heavier resonances. The $\U(1)$ gauge group is expected to be in the Higgs phase if the theory is chiral and the vacuum is aligned in the $G_{\rm SM}$ preserving direction. The dynamically generated masses for the light states can be computed in a systematic way by including the weak gauging of $\U(1)\times G_{\rm SM}$ in a chiral lagrangian approach for the low energy theory of $G$, as discussed in detail in Ref.~\cite{Contino:2020god}. The conditions for anomaly cancellation and their solutions are in any regard independent from the previous considerations on the infrared dynamics of the theories.

We shall be interested in the conditions for the cancellation of local gauge anomalies for $G \times \U(1) \times G_{\rm SM}$. Global (non-perturbative) anomalies~\cite{Witten:1982fp,Witten:1985xe} are related to the topology of the gauge group and depend on the specific choice of Lie group rather than just its Lie algebra. Their analysis is left to a future work (see Refs.~\cite{Garcia-Etxebarria:2018ajm,Davighi:2019rcd,Wan:2020ynf} for a recent study in the SM and some of its extensions).\\
Since the generators of a representation of a semisimple Lie algebra are traceless, the mixed anomalies  $G \times [\U(1)]^2$, $G \times [G_{\rm SM}]^2$, $G_{\rm SM} \times [\U(1)]^2$ and $G_{\rm SM} \times [G]^2$  vanish trivially. Moreover, the $[G_{\rm SM}]^3$ anomaly vanishes as a consequence of the cancellation of gauge anomalies in the SM (or its semisimple extension) and the fact that the set of additional fermions we introduced is vectorlike with respect to $G_{\rm SM}$. Similarly, the anomaly $[G]^3$ vanishes due to the vectorlike structure of the new fermions with respect to $G$.
We are left with the mixed anomalies $\U(1) \times [G_{\rm SM}]^2$, $\U(1) \times [G]^2$, and the Abelian anomaly $[\U(1)]^3$. The only nontrivial condition from the cancellation of local gravitational anomalies is $U(1) \times [\rm grav]^2$, but in the class of models we consider the latter is equivalent to the condition for $\U(1) \times [G]^2$. We thus obtain a system of three equations of the form:
\begin{subnumcases}{\label{eq:general_system}}
\displaystyle \sum_{i=1}^{n_f} (p_{i}+q_{i}) T_i = 0,\label{eq:general_system_1}\\
\displaystyle \sum_{i=1}^{n_f} (p_{i}+q_{i})d_i = 0,\label{eq:general_system_2}\\
\displaystyle \sum_{i=1}^{n_f} (p_{i}^{3}+q_{i}^{3})d_i = 0,\label{eq:general_system_3}
\end{subnumcases}
where we have dropped the common factors $T^{(G)}$ and $d^{(G)}$, which are the same for all fermions and always nonzero.
With our notation and conventions this is a system of Diophantine equations with integer coefficients for the integer charges $(q_i,p_i)$. Since the equations are defined by homogeneous polynomials, it is sufficient to work over the field $\mathbb{Q}$ of rational numbers, and then rescale by an integer multiple to obtain all the integer solutions.

\section{General solution to the anomaly cancellation conditions}\label{sec:3}

In this section we focus on theories in which either the new fermions $(\psi_i,\chi_i)$ have vanishing hypercharge or the SM gauge group is extended to an arbitrary compact semisimple Lie group that we denote by $G_{\rm SM}$. We wish to provide a general solution to the system (\ref{eq:general_system}) in the case of $n_f$ distinct $\U(1)$ charges for the fields $\psi_i$ and $n_f$ for the fields $\chi_i$. For $n_f=2,3$ the cubic equation can often be reduced to a quadratic equation, and we shall treat these cases first. The general case will then be addressed, providing a solution valid also in the previous instances, but corresponding to a different parametrisation. 

\subsection{Models with $n_{f}=2$}
For $n_f=2$, the two linear equations 
\begin{equation}\label{eq:lin}
(p_1+q_1)T_1 +(p_2+q_2)T_2 = 0, \quad \quad (p_1+q_1)d_1 +(p_2+q_2)d_2=0,
\end{equation}
admit solutions with $(p_i+q_i)\neq 0$ only if $T_1 d_2 = d_1 T_2$, in which case they become equivalent. 

In the case $d_1= d_2$ the system reduces to that of four $\U(1)$ charges with zero sum and vanishing sum of cubes. The general solution is well known and is given by assignments of the form 
\begin{equation}\label{eq:vl}
(q_1 = -p_1, q_2 = -p_2), \quad \,\,\quad (p_2 = -p_1, q_2 = -q_1), \quad \text{and} \quad  (p_2 = -q_1, q_2 = -p_1).
\end{equation}
These assignments are vectorlike with respect to $U(1)$, but can still give rise to overall chiral models, which fall into three classes and have been classified in Ref.~\cite{Contino:2020god}. 
From now on we therefore assume without loss of generality that $d_2 > d_1$.

The linear equations \eqref{eq:lin} and the cubic equation
\begin{equation}
(p_1^3+q_1^3)d_1 +(p_2^3+q_2^3)d_2=0,
\end{equation}
always admit a vectorlike set of solution with $(q_1 = -p_1, q_2 = -p_2)$. On the other hand, if $(p_1+q_1)\neq0$, combining the equations and after straightforward manipulations one arrives at
\begin{equation}\label{eq:nf2}
(d_2^2-d_1^2) (p_1+q_1)^2+3d_2^2 (p_1-q_1)^2-3d_2^2(p_2-q_2)^2 =0.
\end{equation}
This is a homogeneous quadratic Diophantine equation in the three variables
\begin{equation}
\label{eq:homogeneous_coord}
X=(p_1+q_1), \quad Y=(p_1-q_1),\quad Z=(p_2-q_2).
\end{equation}
Equations of this type have been extensively studied and they are fully understood, see for instance~\cite{Mordell:1969}.
In particular there are general criteria for the solvability of these equations and it is known that if a particular nontrivial solution is known then there are actually infinite solutions, and the general solution can be parametrised in closed form in terms of arbitrary integers (Theorem 4 in Chapter 7 of~\cite{Mordell:1969}). Let us give a brief description of how this can be achieved in the explicit example of our equation. 

From a geometric point of view, eq.~\eqref{eq:nf2} defines a conic in the projective plane $\mathbb{PR}^2$, which in homogeneous coordinates $[X:Y:Z]$ is given by the zero locus of the quadratic form
\begin{equation}
Q(X,Y,Z) \equiv (d_2^2-d_1^2)\,  X^2+3 d_2^2\, Y^2 -3 d_2^2\, Z^2.
\end{equation} 
The solutions to the Diophantine equation will be given by all the rational points on this conic, and define a curve in the projective space $\mathbb{PQ}^2$. Given a rational point $P_0 = [X_0:Y_0:Z_0]$ on the conic, any other point on the curve is rational if and only if it lies on a rational line\footnote{\emph{i.e.} a line with rational coefficients.} passing through $P_0$.  Indeed, given any other rational point $\bar{P}$ on the conic there exists a line passing through $\bar{P}$ and $P_0$ and this line has rational coefficients; vice versa, the intersection of a quadratic curve with a rational line through one of its rational point is always a rational point. In this way, all solutions can be generated from a single one.

In practice, it is convenient to eliminate the redundancy under coordinate rescalings and work in affine space. Since $d_2>d_1$ there are no nontrivial solutions with $Z=0$, and we are free to divide by $Z$ to switch to affine coordinates $x=X/Z$ and $y=Y/Z$ and consider the quadratic form
\begin{equation}
q(x,y) \equiv Q\left(\frac{X}{Z},\frac{Y}{Z},1\right) = (d_2^2-d_1^2)\,  x^2+3 d_2^2\, y^2 -3 d_2^2.
\end{equation}
The zero set of $q(x,y)$ defines an ellipse in $\mathbb{R}^2$, which corresponds to our conic. A particular rational point on the ellipse can easily be found by inspection and is given by $(x_0,y_0)=(0,1)$.  All the other rational points will be given by the intersection of the curve with the set of affine lines through $(x_0,y_0)$ with rational coefficients: $y = k \,x +1$ with $k \in \mathbb{Q}$. In particular, by carrying out the algebra it follows that such a point is given by:
\begin{equation}
x =\frac{f_x(k)}{f(k)} \equiv \frac{6 k d_2^2}{d_1^2-d_2^2(3k^2+1)}, \quad \quad y =\frac{f_y(k)}{f(k)} \equiv \frac{d_1^2+d_2^2(3k^2-1)}{d_1^2-d_2^2(3k^2+1)}, \quad \quad f(k)\equiv d_1^2-d_2^2(3k^2+1),
\end{equation}
which is always a rational point as expected.
To go back to the homogeneous space of integer solutions, one can simultaneously rescale all coordinates by $f(k)$ and homogenise the polynomials:
\begin{equation}
X = \ell^2 f_x(k/ \ell), \quad \quad  \quad  Y= \ell^2 f_y(k/ \ell),   \quad \quad \quad Z= \ell^2 f (k/ \ell).
\end{equation}
Since we have assumed $X=(p_1+q_1)\neq 0$ in deriving the quadratic equation we need to exclude the values $k=0$ or $\ell=0$ in order to avoid a double counting.

Finally, the general solution for the $U(1)$ charges can be recovered by use of eqs.~\eqref{eq:lin} and~\eqref{eq:homogeneous_coord}:
\begin{equation}\label{eq:gn2}
\begin{split}
& p_1= \frac{n}{\mu_2}\tilde{p}_1 = \frac{n}{\mu_2} \Big[ d_1^2 \ell^2 +d_2^2(3k^2+6k \ell-\ell^2) \Big], \\
& q_1 = \frac{n}{\mu_2}\tilde{q}_1 = \frac{n}{\mu_2} \Big[-d_1^2 \ell^2 +d_2^2(\ell^2 +6k \ell-3k^2) \Big],\\
& p_2 = \frac{n}{\mu_2}\tilde{p}_2 = \frac{n}{\mu_2} \Big[d_1^2 \ell^2-6 d_1d_2 k \ell -d_2^2(3k^2+\ell^2)\Big], \\
&q_2= \frac{n}{\mu_2}\tilde{q}_2 = \frac{n}{\mu_2} \Big[-d_1^2 \ell^2-6 d_1d_2 k  \ell +d_2^2(3k^2+\ell^2)\Big],
\end{split}
\end{equation}
where $n,k,\ell \in \mathbb{Z} \setminus \{0\}$, the $\{ \tilde{p}_i, \tilde{q}_i \}$ are defined implicitly above, and
\begin{equation}
\mu_2 = {\rm{gcd}} \Big( \tilde{p}_1,\tilde{p}_2,\tilde{q}_1,\tilde{q}_2 \Big),
\end{equation}
plus the vectorlike set of solutions with $(q_1 = -p_1, q_2 = -p_2)$. The factor $n$ takes into account the possibility of rescaling a solution by an arbitrary integer. 

\subsection{Models with $n_{f}=3$}\label{ssc:n3}

Increasing the number of independent charges we consider now theories with $n_{f}=3$.
The system of equations comprises the two linear equations 
\begin{equation}\label{eq:lin_3}
(p_1+q_1)T_1 +(p_2+q_2)T_2 +(p_3+q_3)T_3 = 0, \quad \quad (p_1+q_1)d_1 +(p_2+q_2)d_2+ (p_3+q_3)d_3=0,
\end{equation}
and the cubic equation
\begin{equation}\label{eq:cub_3}
(p_1^3+q_1^3)d_1+(p_2^3+q_2^3)d_2+(p_3^3+q_3^3)d_3 =0.
\end{equation} 
This system always admits vectorlike solutions with $(p_i+q_i)= 0$ for $i=1,2,3$. Moreover, if there is at least a vectorlike pair of fermions then the equations reduce to those of the $n_f =2$ case. We therefore restrict our attention to the case $(p_i+q_i)\neq 0$.

For notational simplicity, we introduce the antisymmetric combinations
\begin{equation}\label{eq:dij}
D_{ij} = -D_{ji} \equiv d_i T_j -d_j T_i,
\end{equation}
which satisfy the cyclic property 
\begin{equation}\label{eq:cyc}
d_1 D_{23}+d_2 D_{31}+ d_3 D_{12}=0.
\end{equation}
We note that if one $D_{ij}= 0$ but the other two $D_{ij}$'s are nonzero, one can easily prove by combining the two linear equations~\eqref{eq:lin_3} that the pair of charges with $(k\neq i,j)$ is vectorlike: $(p_k+q_k)= 0$. In this case, therefore, the equations reduce to those of the $n_f =2$ case we already treated.

\subsubsection{Nondegenerate system}

Let us consider the case $D_{ij}\neq 0$ for every $i,j$. From eq.~\eqref{eq:cyc} it follows that the signs of the $D_{ij}$'s cannot all be equal, and with a suitable permutation of the indices one can always assume ${\rm{sgn}} \big( D_{23} \big) = - {\rm{sgn}} \big( D_{12} \big)= -{\rm{sgn}} \big( D_{31} \big)$. We shall assume that such a reshuffling has been performed and that $D_{23}$ is the one with opposite sign.

In this case the system in \eqref{eq:general_system} can be reduced to a homogeneous quadratic equation and so the general procedure to solve the system parallels that of $n_f=2$, although the computations become more involved. From the two linear equations~\eqref{eq:lin_3} one can easily eliminate two variables as
\begin{equation}\label{eq:slin}
p_2 +q_2 = \frac{D_{31}}{D_{23}}\,  (p_1+q_1),  \quad \quad \quad
p_3+q_3 = \frac{D_{12}}{D_{23}} \, (p_1+q_1).
\end{equation}
Then, direct substitution allows one to turn the cubic equation~\eqref{eq:cub_3} into a quadratic one
\begin{equation}
 \big[ d_1 D_{23}^3 +d_2 D_{31}^3+d_3 D_{12}^3 \big] \, X^2 
 +3  D_{23}^2  \big[  d_1 D_{23} \,W^2+d_2 D_{31}\, Y^2 +d_3  D_{12}\, Z^2 \big]
= 0,
\end{equation}
in terms of the four variables
\begin{equation}
\label{eq:homogeneous_coord2}
X=(p_1+q_1), \quad W=(p_1-q_1), \quad Y=(p_2-q_2),\quad Z=(p_3-q_3),
\end{equation}
which can be solved using the same method described previously. 

Analogously to the $n_f =2 $ case, solutions can be expressed as the rational points of a hypersurface in the projective space $\mathbb{PR}^3$, that in homogeneous coordinates $[X:Y:Z:W]$ is defined by the zero locus of the quadratic form
\begin{equation}
 Q(X,Y,Z,W) \equiv \, X^2 \big[ d_1 D_{23}^3 +d_2 D_{31}^3+d_3 D_{12}^3 \big] 
+3  D_{23}^2 \big[ d_1 D_{23} W^2+  d_2 D_{31} Y^2+d_3  D_{12}Z^2  \big].
\end{equation}
They thus define a quadric hypersurface in the rational projective space  $\mathbb{PQ}^3$. It is convenient to switch again to affine coordinates  $x = X/W, \, y = Y/W, z = Z/W$, where we are neglecting for the moment the solutions with $W=0$, which correspond to points at infinity in the affine space spanned by $(x,y,z)$.\footnote{Notice that we are choosing to go to affine space by rescaling  by $W$ in order to minimise the number of points at infinity of the quadric, since $D_{23}$ was the coefficient with opposite sign with respect to $D_{12},D_{13}$.} The quadratic form becomes
\begin{equation}
\begin{split}
 q(x,y,z) \equiv  \, & Q(X/W,Y/W,Z/W,1) = \\
 & x^2 \big[ d_1 D_{23}^3 +d_2 D_{31}^3+d_3 D_{12}^3 \big] 
+3  D_{23}^2 \big[   d_1 D_{23}+y^2 d_2 D_{31}+z^2 d_3  D_{12} \big] .
\end{split}
\end{equation}
The rational point $(x_0,y_0,z_0)=(0,1,1)$ always belongs to the surface, since the $D_{ij}$'s satisfy the cyclic property (\ref{eq:cyc}). From the knowledge of a rational solution, all the other rational points can be obtained as the intersection between the surface and the set of rational lines through $(x_0,y_0,z_0)$, defined by $$y = k\,x +1,\quad z= \ell \, x + 1,$$ for $k, \ell \in \mathbb{Q}$. 

After a coordinate rescaling and homogenisation, the full solution for the $\rm{U}(1)$ charges can finally be expressed as
\begin{equation}
\begin{split}
p_1 =\frac{n}{\mu_3} \tilde{p}_1 = \frac{n}{\mu_3} \Big( &\, m^2 \big[d_1 D_{23}^3 + d_2 D_{31}^3 +d_3 D_{12}^3\big]  
  -\,6 m D_{23}^2 \big [ k\, d_2 D_{31}+ \ell\, d_3D_{12} \big] \,\\
  &  + D_{23}^2 \big[3  \ell^2 \, d_3 D_{12} + 3  k^2\, d_2 D_{31}  \big]\Big),\\
  q_1 =\frac{n}{\mu_3} \tilde{q}_1 = \frac{n}{\mu_3}\Big( &- m^2 \big[d_1 D_{23}^3 + d_2 D_{31}^3 +d_3 D_{12}^3\big]  
  -\,6 m D_{23}^2 \big [ k\, d_2 D_{31}+ \ell\, d_3D_{12} \big] \,\\
  &  - D_{23}^2 \big[3  \ell^2 \, d_3 D_{12} + 3  k^2\, d_2 D_{31}  \big] \Big),\\
p_2 =\frac{n}{\mu_3} \tilde{p}_2 = \frac{n}{\mu_3} \Big( &\,  m^2 \big[d_1 D_{23}^3 + d_2 D_{31}^3 +d_3 D_{12}^3\big]  
- 6 m D_{31} D_{23} \big[
k \, d_2 D_{31}+\ell \, d_3D_{12}  \big]  \\
& -6 k D_{23}^2 \big[ k\, d_2 D_{31} + \ell \,d_3 D_{12} \big]   + D_{23}^2 \big[3  \ell^2 \, d_3D_{12} + 3  k^2\, d_2 D_{31} \big] \Big), \\
q_2 =\frac{n}{\mu_3} \tilde{q}_2 = \frac{n}{\mu_3} \Big( &- m^2 \big[d_1 D_{23}^3 + d_2 D_{31}^3 +d_3 D_{12}^3\big]  
- 6 m D_{31} D_{23} \big[
k \, d_2 D_{31}+\ell \, d_3D_{12}  \big]  \\
& +6 k D_{23}^2 \big[ k\, d_2 D_{31} + \ell \,d_3 D_{12} \big]  - D_{23}^2 \big[3  \ell^2 \, d_3D_{12} + 3  k^2\, d_2 D_{31} \big]  \Big),\\
p_3 =\frac{n}{\mu_3} \tilde{p}_3 = \frac{n}{\mu_3} \Big( & \,  m^2 \big[d_1 D_{23}^3 + d_2 D_{31}^3 +d_3 D_{12}^3\big]  
- 6 m D_{12} D_{23}\big[ k\, d_2 D_{31}+ \ell\, d_3D_{12}\big] \\
& - 6 \ell D_{23}^2 \big[ k\, d_2 D_{31}+ \ell\, d_3D_{12}\big]  + D_{23}^2 \big[ 3  \ell^2 \, d_3D_{12} + 3  k^2\, d_2 D_{31} \big] \Big), \\
q_3 =\frac{n}{\mu_3} \tilde{q}_3 = \frac{n}{\mu_3} \Big( & - m^2 \big[d_1 D_{23}^3 + d_2 D_{31}^3 +d_3 D_{12}^3\big]  
- 6 m D_{12} D_{23}\big[ k\, d_2 D_{31}+ \ell\, d_3D_{12}\big] \\
& + 6 \ell D_{23}^2 \big[ k\, d_2 D_{31}+ \ell\, d_3D_{12}\big]  - D_{23}^2 \big[ 3  \ell^2 \, d_3D_{12} + 3  k^2\, d_2 D_{31} \big] \Big),
\end{split}
\end{equation}
with $k,\ell, m,n \in \mathbb{Z}$ and the $\{ \tilde{p}_i, \tilde{q}_i \}$ are defined implicitly above.
Again, the solutions are parametrised by an extra variable $n$ which corresponds to an overall rescaling. The normalisation constant $\mu_3$ is given by
\begin{equation}
\mu_3 = {\rm{gcd}} \Big( \tilde{p}_1,\tilde{p}_2,\tilde{p}_3,\tilde{q}_1,\tilde{q}_2,\tilde{q}_3 \Big).
\end{equation}

The last question left to determine is whether the points at infinity, corresponding to $W = 0$, give rise to an additional family of solutions. In this case, the equation reduces to
\begin{equation}\label{eq:W0}
\, X^2 \big[ d_1 D_{23}^3 +d_2 D_{31}^3+d_3 D_{12}^3 \big] 
+3  D_{23}^2 \big[ Y^2 d_2 D_{31}+Z^2 d_3  D_{12} \big] =0.
\end{equation}
If the sign of the $X^2$ coefficient ${\rm{sgn}} \big(d_1 D_{23}^3 +d_2 D_{31}^3+d_3 D_{12}^3 \big)$ is equal to  ${\rm{sgn}} \big(D_{12} \big) =
{\rm{sgn}} \big(D_{31} \big)$, eq. (\ref{eq:W0}) does not admit any nontrivial solutions. If the signs are different, the (non) existence of additional solutions depends on the numerical value of the coefficients. A known result in Number Theory (See Chapter 7 of \cite{Mordell:1969}, Theorem 5) guarantees that if an equation of the form
\begin{equation}\label{eq:pt}
a\, X^2 +b\,Y^2+c\, Z^2 = 0, \quad \quad \quad a \in \mathbb{Z} \quad b,c  \in \mathbb{N} \setminus \{ 0\},
\end{equation}
has nontrivial integer solutions, at least one of them satisfies
\begin{equation}\label{eq:int}
|X_0| \leq \sqrt{b c}, \quad \quad\,|Y_0| \leq \sqrt{|a| c}, \quad \quad \,|Z_0| \leq \sqrt{|a|b}.
\end{equation} For any choice of representations, it is therefore possible to determine whether (\ref{eq:W0}) is solvable in a finite number of steps, by checking whether any of the integer triples satisfying (\ref{eq:int}) solve the equation. In the affirmative case, all the other solutions can be generated from a particular one with the methods employed in the last two sections. All three possibilities can be realised in practice, as shown by the examples in Appendix \ref{app:A}.

\subsubsection{Degenerate system}

If $D_{ij} =0$ for every $(i,j)$ then the two linear equations~\eqref{eq:lin_3} become equivalent. In this case the remaining equations correspond to a special instance of the equations for the cancellation of gauge anomalies for a single $U(1)$ gauge group studied in Refs.~\cite{Costa:2019zzy,Allanach:2019gwp}. In particular, they coincide with the anomaly cancellation equations for a set of $2 d_1 + 2 d_2 + 2 d_3$ fermions charged under a single $U(1)$ gauge group, in which there are $d_i$ pairs with charges $(p_i,q_i)$ for $i=1,2,3$. The method of solution of Ref.~\cite{Costa:2019zzy} does not allow one to easily enforce the constraint that some of the charges are equal. However it is still possible to obtain a closed form solution following an observation stressed in Appendix~B of Ref.~\cite{Allanach:2019gwp}: given a cubic curve with a known rational point $P_0$ which is a double point\footnote{That is a zero of the cubic and of all its partial derivatives.}, every other rational point of the curve can be expressed in terms of a rational parameter (see \emph{e.g.} Theorem 3 in Chapter 9 of~\cite{Mordell:1969}). This is a generalisation of the method we used in the case of a quadric. It is sufficient to consider the set of all the rational lines through $P_0$: this set covers the whole rational projective space of interest and the intersection of such a line with the cubic is always a rational point, since $P_0$ is a double point.

In our case we can use the linear equation~\eqref{eq:lin_3} to eliminate $q_3$ and obtain a cubic in $\mathbb{PQ}^4$, defined in terms of homogeneous coordinates $[p_1:q_1:p_2:q_2:p_3]$ by the zero locus of
\begin{equation}\label{eq:cubic_sol}
F\left(p_1,q_1,p_2,q_2,p_3\right) = d_1 d_3^2 \, (p_1^3+q_1^3)+ d_2 d_3^2 \, (p_2^3+q_2^3) + d_3^3 \, p_3^3 - \big(d_1 (p_1+q_1) + d_2 (p_2+q_2) + d_3\, p_3 \big)^3 .
\end{equation} 
It is easy to verify that the rational point $\Pi_0=[1:-1:1:-1:1]$ lies on the cubic and is also a double point, since all the partial derivatives of $F$ vanish. We can therefore find all the rational points of the cubic $F$ as the intersection of $F$ with the rational lines
\begin{equation}\label{eq:rational_lines}
L = k_1 \Pi_0 + k_2 \Sigma,
\end{equation}
where $K = [k_1 : k_2] \in \mathbb{PQ}$ and $\Sigma = [\ell_1: m_1 : \ell_2 : m_2 : \ell_3] \in \mathbb{PQ}^4$ (the notation for the components of $\Sigma$ is chosen for convenience, in analogy with the notation we use for the charges).
Substituting \eqref{eq:rational_lines} in \eqref{eq:cubic_sol} we find the intersection points as the solutions of
\begin{equation}
k_2^2 \Bigg(k_1 A_1 + k_2 A_2 \Bigg)=0,
\end{equation}
where
\begin{equation}\label{eq:As}
\begin{split}
&A_1 = 3 d_1 d_3^2 ( \ell_1^2 - m_1^2) +  3 d_2 d_3^2 ( \ell_2^2 - m_2^2) +  3 d_3^3 \, \ell_3^2 
 - 3  d_3 \big(d_1 (\ell_1+m_1) + d_2 (\ell_2+m_2) + d_3\, \ell_3 \big)^2,\\
&A_2 = d_1 d_3^2 ( \ell_1^3 + m_1^3) +  d_2 d_3^2 ( \ell_1^3 + m_1^3)+
 d_3^3 \, \ell_3^3 
 - \big(d_1 (\ell_1+m_1) + d_2 (\ell_2+m_2) + d_3\, \ell_3 \big)^3.
\end{split}
\end{equation}
For $k_2=0$ we recover the original point $\Pi_0$, whereas for $[k_1:k_2]=[A_2:-A_1]$ we find all the other rational points on the cubic.
Plugging back and using the linear equation to recover $q_3$, we obtain the general solution for the charges:
\begin{equation}\label{eq:sol_degenerate_3}
\begin{split}
& p_1= \frac{n}{\mu_3}\tilde{p}_1 = \frac{n}{\mu_3} \Big[ A_2 - A_1 \, \ell_1 \Big], \\
& q_1 = \frac{n}{\mu_3}\tilde{q}_1 = \frac{n}{\mu_3} \Big[ -A_2 - A_1 \, m_1 \Big], \\
& p_2 = \frac{n}{\mu_3}\tilde{p}_2 = \frac{n}{\mu_3} \Big[ A_2 - A_1 \, \ell_2 \Big], \\
& q_2= \frac{n}{\mu_3}\tilde{q}_2 = \frac{n}{\mu_3} \Big[ -A_2 - A_1 \, m_2 \Big], \\
& p_3 = \frac{n}{\mu_3}\tilde{p}_3 = \frac{n}{\mu_3} \Big[ A_2 - A_1 \, \ell_3 \Big], \\
& q_3= \frac{n}{\mu_3}\tilde{q}_3 = \frac{n}{\mu_3} \Big[- A_2 + \dfrac{A_1}{d_3}\Big(d_1 \, (\ell_1+m_1) + d_2 (\ell_2+m_2) + d_3 \, \ell_3 \Big)\Big], \\
\end{split}
\end{equation}
with $A_1,A_2$ defined in~\eqref{eq:As}, $\ell_i,m_i, n\in \mathbb{Z}$ and the $\{ \tilde{p}_i, \tilde{q}_i \}$ defined implicitly above.
As before, the integer solutions are parametrised by an extra variable $n$ which corresponds to an overall rescaling, and the normalisation constant $\mu_3$ is defined as
\begin{equation}
\mu_3 = {\rm{gcd}} \Big( \tilde{p}_1,\tilde{p}_2,\tilde{p}_3,\tilde{q}_1,\tilde{q}_2,\tilde{q}_3 \Big).
\end{equation}
Notice that $A_1$ is a multiple of $d_3$ and so the $\{ \tilde{p}_i, \tilde{q}_i \}$ are manifestly integers. Moreover the vectorlike solutions are recovered when $A_1 =0$.

\subsection{Models with arbitrary $n_f$}\label{sec:gen_sol}

In this section we set $\nu \equiv n_f$ for notational convenience.
The method of solution described in the last paragraph can be straightforwardly extended to treat the general case. Let us consider first the equations~\eqref{eq:general_system_2} and~\eqref{eq:general_system_3}, involving the dimensions $d_i$. We can use the linear equation to solve for $q_\nu$ and plug back in~\eqref{eq:general_system_3} to obtain a cubic in $\mathbb{PQ}^{2\nu-2}$. In terms of homogeneous coordinates $[p_1:q_1:\dots:p_{\nu-1}:q_{\nu-1}:p_\nu]$ it is defined by
\begin{equation}\label{eq:cubic_sol_nf}
F\left(p_i,q_i\right) = d_{\nu}^2 \sum_{i=1}^{\nu-1} d_i  \, (p_i^3+q_i^3)+ d_\nu^3 \, p_\nu^3 - \left(\sum_{i=1}^{\nu-1} d_i ( p_i + q_i)+ d_{\nu}\, p_{\nu} \right)^3 .
\end{equation} 
It is easy to check that it describes a singular cubic hypersurface with $\Pi_0=[1:-1:\dots:1:-1:1]$ as a rational double point.
As previously described, we can find  all the rational points of the cubic $F$ as the intersection of $F$ with the rational lines
\begin{equation}\label{eq:rational_lines_nf}
L = k_1 \Pi_0 + k_2 \Sigma,
\end{equation}
where $K = [k_1 : k_2] \in \mathbb{PQ}$ and $\Sigma = [\ell_1: m_1 :\dots: \ell_{\nu-1}: m_{\nu-1} : \ell_\nu] \in \mathbb{PQ}^{2\nu-2}$.

Carrying out the algebra we find that the rational points are given by
\begin{equation}\label{eq:sol_general}
\begin{split}
& p_i= \frac{n}{\mu_{\nu}}\tilde{p}_1 = \frac{n}{\mu_{\nu}} \Big[ A_2 - A_1 \, \ell_i \Big], \\
& q_i = \frac{n}{\mu_{\nu}}\tilde{q}_1 = \frac{n}{\mu_{\nu}} \Big[ -A_2 - A_1 \, m_i \Big], \\
& p_{\nu} = \frac{n}{\mu_{\nu}}\tilde{p}_\nu = \frac{n}{\mu_{\nu}} \Big[ A_2 - A_1 \, \ell_{\nu} \Big], \\
& q_{\nu}= \frac{n}{\mu_{\nu}}\tilde{q}_\nu = \frac{n}{\mu_{\nu}} \Big[- A_2 + \dfrac{A_1}{d_{\nu}}\left(\sum_{i=1}^{\nu-1} d_i ( \ell_i + m_i)+ d_{\nu}\, \ell_{\nu} \right)\Big], \\
\end{split}
\end{equation}
where $\ell_i,m_i, n\in \mathbb{Z}$, the index $i$ runs on $i=1, \dots , \nu-1$ and $\mu_\nu$, $\{ \tilde{p}_i, \tilde{q}_i \}$ are defined in analogy to the previous section.
The coefficients $A_1,A_2$ are polynomials in the parameters $\ell_i,m_i$:
\begin{equation}\label{eq:As_general}
\begin{split}
&A_1 = 3 d_{\nu}^2 \sum_{i=1}^{\nu-1} d_i ( \ell_i^2 - m_i^2) +   3 d_{\nu}^3  \, \ell_{\nu}^2 
 - 3  d_{\nu} \left(\sum_{i=1}^{\nu-1} d_i ( \ell_i + m_i)+ d_{\nu}\, \ell_{\nu} \right)^2,\\
&A_2 = d_{\nu}^2 \sum_{i=1}^{\nu-1} d_i ( \ell_i^3 + m_i^3)+
 d_{\nu}^3\,  \ell_3^3
 - \left(\sum_{i=1}^{\nu-1} d_i ( \ell_i + m_i)+ d_{\nu}\, \ell_{\nu} \right)^3.
\end{split}
\end{equation}
The linear equation involving the Dynkin indices~\eqref{eq:general_system_1} results in an additional constraint on the arbitrary integers $\ell_i,m_i$:
\begin{equation}
\sum_{i=1}^{\nu-1} D_{i\nu} (\ell_i + m_i) =0,
\end{equation}
which can be either solved explicitly in the previous expressions or, more conveniently, taken into account when choosing the integers parametrising the solution.

\section{The inclusion of hypercharge}
\label{sec:hypercharge}

Having found the general charge assignment for the $U(1)$ gauge group in cases where $G_{\rm SM}$ is a semisimple extension of the SM group we wish now to address the case in which $G_{\rm SM}$ is taken to be  $\SU(3)_{\rm QCD}\times \SU(2)_{\rm EW}\times \U(1)_Y$, and the new fermions $\psi_i,\chi_i$ have nonvanishing hypercharge.
The quantum numbers under $G \times U(1) \times G_{\rm SM} $ are therefore
given by (\ref{eq:genmod}), with the specification $r_i = \sum_\alpha (\hat{r}_{i,\alpha},y_{i,\alpha})$, where $\hat{r}_{i,\alpha}$ is an irreducible representation of $\SU(3)_{\rm QCD}\times \SU(2)_{\rm EW}$, and $y_{i,\alpha}$ their corresponding hypercharge. The anomaly cancellation conditions for the $U(1)$ charges, and their solutions discussed in the previous section, are still relevant for the models we consider here, but must be supplemented by additional conditions. There are only two additional nontrivial equations for the cancellation of gauge anomalies, namely the mixed ones $\U(1)_Y \times [\U(1)]^2$ and$\U(1) \times [\U(1)_Y]^2$. 
In the general case they read
\begin{equation}\label{eq:general_y}
\begin{cases}
\displaystyle \sum_{i=1}^{n_f} (p_i+q_i)\sum_\alpha d_{i,\alpha} y_{i,\alpha}^2 =0, \\
\displaystyle \sum_{i=1}^{n_f} (p_i^2-q_i^2) \sum_\alpha d_{i,\alpha} y_{i,\alpha} = 0.
\end{cases}
\end{equation}
The approach that we choose to pursue is the following: given a choice of $U(1)$ charges $(p_i,q_i)$, which for fixed $\SU(3)_{\rm QCD}\times \SU(2)_{\rm EW}$ representations can be found using the results of the previous section, we want to determine which assignments of hypercharge are consistent with the system~\eqref{eq:general_y}.

It is always possible to solve the system by solving for one of the hypercharges in the linear equation and plugging back in the quadratic one. We are left with an homogeneous quadratic Diophantine whose solutions can always be found with the methods already illustrated. We shall not do this explicitly in the general case, but the procedure to find the general solution for any case of interest should be clear. We present, however, the explicit solution in two special cases of phenomenological relevance, in connection to models of composite axions (see Section~\ref{sec:composite_axions} for a brief overview).

\subsection{Models with $n_{f}=2$}\label{sec:hyper_nf2}

We treat the case in which $y_{i,\alpha}= y_{i}$ for every $\alpha$.
The equations for the hypercharges read:
\begin{equation}\label{eq:ny}
(p_1+q_1)d_1 y_1^2+(p_2+q_2)d_2 y_2^2 =0 \quad \quad \quad 
(p_1^2-q_1^2)d_1 y_1+(p_2^2-q_2^2)d_2 y_2 = 0.
\end{equation}
In the case of vectorlike assignment $(q_1 = -p_1, q_2 = -p_2)$ they are trivially satisfied for arbitrary $y_1,y_2$. We shall therefore consider the case $(p_1+q_1)\neq0$.

If $d_1=d_2$, the $\U(1)$ charges satisfy one of the assignments in \eqref{eq:vl}. It follows that the general solutions with $(p_1+q_1)\neq0$ are:  
\begin{equation}
(p_2 = -p_1, q_2 = -q_1, y_2 = -y_1), \qquad (p_2 = -q_1, q_2 = -p_1, y_2 = y_1).
\end{equation}

Let us now consider $d_2 > d_1$. Then, the first equation in \eqref{eq:ny} (in combination with eq.~\eqref{eq:lin}), implies that $y_1 = \pm y_2$. Because of the second equation in \eqref{eq:ny}, however, these assignments are consistent only if the $\U(1)$ charges satisfy respectively
\begin{equation}\label{eq:n2y}
 Q^+\equiv (p_1^2-q_1^2)d_1 + (p_2^2-q_2^2)d_2 = 0, \quad \quad \text{or} \quad \quad   Q^- \equiv (p_1^2-q_1^2)d_1 - (p_2^2-q_2^2)d_2 = 0.
\end{equation}
Since the possible values that $\{ p_i,q_i\}$ can take are already determined by (\ref{eq:gn2}), it is possible to verify whether any of them satisfy \eqref{eq:n2y} upon direct substitution. After a few algebraic manipulations,
\begin{equation}
Q^+ = 144 d_1 d_2^4 \,k^3 \ell ,\quad \quad \quad Q^- =48 d_1 (d_1^2 - d_2^2) d_2^2 \,k \ell^3 .
\end{equation}
The only solutions are either $k=0$ or $\ell=0$, which are already included in the solutions with $(p_1+q_1) = 0$. We conclude that for $d_2 > d_1$ the only consistent assignments of hypercharge with $y_{i,\alpha}= y_{i}$ are the vectorlike ones with $(q_1 = -p_1, q_2 = -p_2)$ and arbitrary $y_1,y_2$.

\subsection{Models with $n_{f}=3$ and irreducible representations}

For $n_{f}=3$ and $r_i$ irreducible the anomaly cancellation conditions take the form
\begin{equation}\label{eq:ny3}
(p_1+q_1)d_1 y_1^2+(p_2+q_2)d_2 y_2^2 + (p_3+q_2)d_3 y_3^2 =0,
\end{equation}
and
\begin{equation}\label{eq:ny3b}
(p_1^2-q_1^2)d_1 y_1+(p_2^2-q_2^2)d_2 y_2 +(p_3^2-q_3^2)d_3 y_3= 0.
\end{equation}
As in the $n_{f}=2$ case, vectorlike solutions with ($p_i = -q_i$) automatically satisfy the equations for any choice of hypercharges, so we assume $p_i+q_i  \neq 0$ in the following. Equation \eqref{eq:ny3}, together with the other linear equations \eqref{eq:lin_3}, immediately implies that for any chiral solution
\begin{equation}
\rm{det} \begin{pmatrix}
d_1 & d_2 & d_3\\
T_1 & T_2 & T_3 \\
d_1 y_1^2 & d_2 y_2^2 & d_3 y_3^2\\ 
\end{pmatrix} = 0,
\end{equation}
also equivalent to
\begin{equation}\label{eq:det}
y_1^2 d_1 D_{23}+ y_2^2 d_2 D_{31}+ y_3^2 d_3 D_{12} =0,
\end{equation}
in our notation. 

We consider the case in which $D_{ij}\neq 0$ for all pairings, that is realised in all the $n_f=3$ composite axion models classified in Ref.~\cite{Contino:2021ayn}, see Tab.~\ref{tab:nf3}. Then, eq.~\eqref{eq:ny3b} can be turned into a simpler form
\begin{equation}
(p_1-q_1)d_1 y_1 D_{23}+ (p_2-q_2)d_2 y_2 D_{31}+ (p_3-q_3)d_3 y_3 D_{12} =0,
\end{equation}
from the relations \eqref{eq:slin} involving the sums $p_i+q_i$. 
Assuming that the system does not admit solutions with $p_i+q_i =0$ for every $i=1,2,3$,\footnote{This is equivalent to the condition $ d_1 D_{23}^3+d_2 D_{31}^3+d_3  D_{12}^3\neq 0$, satisfied in all the explicit models of interest.} and without loss of generality taking $p_3\neq q_3$, upon further substitution we obtain a second degree homogeneous equation in two variables for the hypercharges
\begin{equation}
\begin{gathered}
y_1^2 d_1 D_{23} \big[d_3 D_{12} (p_3-q_3)^2 +d_1 D_{23}(p_1-q_1)^2\big]+
y_2^2 d_2 D_{31} \big[d_3 D_{12} (p_3-q_3)^2 +d_2 D_{31}(p_2-q_2)^2\big]\\+ 2 y_1 y_2 d_1 d_2 D_{23}D_{31}(p_1-q_1)(p_2-q_2) = 0,
\end{gathered}
\end{equation}
which can easily be solved for given values of the charges $\{p_i,q_i\}$. In particular, 
it admits a solution if and only if the quantity 
\begin{equation}
\Delta = \sqrt{- d_1 d_2 D_{23}D_{31}(p_3-q_3) \big[ d_1 D_{23} (p_1-q_1)+ d_2 D_{31} (p_2-q_2)+d_1 D_{23}(p_3-q_3)\big]} \in \mathbb{Z}.
\end{equation}
The values of the hypercharges $y_2,y_3$ can be straightforwardly derived from the previous equations given an arbitrary integer value of $y_1$.

\section{Composite axion models and Peccei--Quinn violating operators}\label{sec:composite_axions}

Having determined the general solutions to the anomaly cancellation equations for a broad class of chiral gauge theories, we wish to apply these results to cases of phenomenological relevance, illustrating their usefulness. We shall do so in the context  of the models of composite QCD axion recently introduced in Ref.~\cite{Contino:2021ayn}, which we briefly review. We refer the reader to ~\cite{Contino:2021ayn} for a more detailed discussion, references to the original literature and alternative approaches.

The models considered have the structure described in eq.~\eqref{eq:genmod}, with $G=\rm SU(N)$ and $R=\square$, with $G_{\rm SM}$ taken to be either the SM gauge group or its unified version $\rm SU(5)_{GUT}$, and the $\rm SU(N)$ dynamics assumed to be confining with a confinement scale of the order of the Peccei--Quinn scale $f_{PQ}\gtrsim 4\cdot 10^8 \, \rm GeV$.
A schematic table depicting the matter content of the models is shown in Table \ref{tab:model}. When the set $\{r_{i,\alpha}\}$ of irreducible $G_{\rm SM}$ representations includes fragments with different Dynkin indices, some of the global symmetries are anomalous and one of them can be identified with the PQ symmetry giving rise to the QCD axion. It is spontaneously broken by the confining dynamics and the QCD axion is identified with one of the (pseudo) Nambu--Goldstone bosons. The accidental nature of the symmetry thus provides a natural explanation for its existence, differently from more traditional constructions where the axion is introduced as the phase of a fundamental complex scalar field and the PQ symmetry is postulated by hand~\cite{Dine:1981rt,Zhitnitsky:1980tq,Kim:1979if,Shifman:1979if}. The models can be seen as generalisations of the composite axion proposed in \cite{Kim:1984pt}, with the difference that fermions are now charged under an additional $\U(1)$ factor. Since the axion decay constant $f_a$ will be of order of the confinement scale $\Lambda_{PQ}$, the latter has to take very high values in realistic models, $\Lambda_{PQ} > \, 10^{8} \, \rm{GeV}$. Therefore, the low energy phenomenology is completely determined by the states that remain naturally light, such as (pseudo) Nambu-Goldstone bosons. The effects of the $U(1)$ and $G_{\rm SM}$ weak gaugings can be included systematically, and the dynamics of these states thoroughly analysed, with the use of an effective chiral lagrangian approach. The phenomenological properties of the QCD axion can be well predicted in terms of the UV theory. In the notation of~\cite{Contino:2021ayn}, and using the results of~\cite{GrillidiCortona:2015jxo}, one has: %
\begin{equation}
\begin{split}
&m_{a}= 5.70 (7) \left(\dfrac{10^{12}\, \rm GeV}{f_{a}}\right) \, \rm \mu eV, \\[0.2cm]
& g_{a\gamma\gamma} = \dfrac{\alpha_{em}}{2\pi f_{a}} \left( \dfrac{E}{N}- 1.92 (4) \right), \\[0.2cm]
& c_{p}=-0.47(3), \quad c_{n}=-0.02(3)\, .
\end{split}
\end{equation}
The vanishing of the leading-order UV contributions to the axion-fermion couplings in our model gives a sharp prediction for these quantities, common to all the so-called ``hadronic" axion models.

\begin{table}
\centering
\begin{tabular}{lccc|c}
& $\rm SU(N)$ & $\UoneD$ & $\GSM$ & $\UonePQ$ \\
\cline{1-5} 
\rule{0pt}{2.4ex}$\psi_1$ & $\square$ &  $p_{1}$ & $r_1$& $\alpha_1$ \\
\vdots & \vdots & \vdots & \vdots & \vdots    \\
$\psi_{\nu}$ & $\square$ &  $p_{\nu}$ & $r_{\nu}$ & $\alpha_{\nu}$\\
\cline{1-5}        
\rule{0pt}{2.4ex}$\chi_1$ & $\bar{\square}$ &  $q_{1}$  & $\bar{r}_1$& $\alpha_1$ \\        
\vdots & \vdots & \vdots & \vdots & \vdots    \\
$\chi_{\nu}$ & $\bar{\square}$ &  $q_{\nu}$   & $\bar{r}_{\nu}$ &  $\alpha_{\nu}$\\
\end{tabular}
\caption{\it Schematic structure of the composite axion models with $n_f=\nu$. The representations $r_1,r_2,\dots,r_\nu$ can be reducible, and the $\UoneD$ charges satisfy $(p_i + q_i) \neq 0$ in chiral models.  The PQ transformation, if it exists, is a traceless, axial, global $\rm{U}(1)$ symmetry which commutes exactly with the whole gauge group and is anomaly-free under $\rm SU(N)^2$. See Ref.~\cite{Contino:2021ayn} for a classification of explicit examples.}
\label{tab:model}
\vspace{0.5cm}
\end{table}

Ref.~\cite{Contino:2021ayn} classified all the choices of $G_{\rm SM}$ representations for $n_f=2$ and $n_f=3$ (in the first case with $r_i$ reducible, while in the second case with $r_i$ irreducible) such that: 1) the renormalisable lagrangian has an accidental Peccei--Quinn global symmetry; 2) the dynamics of the $\rm SU(N)$ gauge group is confining; 3) the SM gauge couplings are perturbative up to the Planck scale. 

Generic higher dimensional operators can and will break explicitly the global PQ symmetry. In particular, these operators are expected to be generated by every consistent quantum gravity UV completion~\cite{Banks:2010zn,Harlow:2018tng}, potentially spoiling the QCD axion solution to the strong CP problem. This UV sensitivity is known as the \emph{axion quality problem}.
To quantify the level of protection needed to ensure a robust solution to the strong CP problem irrespectively of the UV physics, it is necessary to identify the leading effects that generate a UV contribution to the QCD axion potential.

It has been shown in Ref.~\cite{Contino:2021ayn} that the PQ violating operators that generate a potential for the QCD axion are gauge invariant operators of the schematic form $\mathcal{O}_{\rm PQ}= \prod_i \psi_{i}\chi_{i}$, possibly nonlocal products of local gauge invariant scalar operators (\emph{i.e.} of the operators that appear in the lagrangian).\footnote{Additional factors of the form $(\psi_{i}^{\dagger}\psi_{i})$, $(\chi_{i}^{\dagger}\chi_{i})$ could in principle be relevant for nonlocal operators, but are always redundant for local operators, since they are always PQ singlets and it is always possible to construct a lower dimensional operator by removing them. We shall neglect them in what follows, and comment on them only when needed.} We shall refer to the insertion of a local operator in a correlator as a \emph{single insertion}, whereas the insertion of a nonlocal operator of the described form will be called a \emph{multiple insertion}. An operator corresponding to a double insertion will be a bilocal operator $\mathcal{O}(x_1,x_2)$, and more general multiple insertions will correspond to multilocal operators.
We assume that the anomalous dimensions of the operators are small corrections to the classical dimension.
The effective dimension of a nonlocal operator built from $N$ local ones is
\begin{equation}\label{eq:smi}
\Delta_{\rm eff} = \sum_{i=1}^N \Delta_i -4(N-1),
\end{equation}
where $\Delta_i $ is the scaling dimension of each local operator, and we are neglecting possible anomalous dimensions. Denoting by $\Delta_{\cancel{PQ}}$ the (effective) dimension of the leading PQ violating effects, a solution of the axion quality problem requires $\Delta_{\cancel{PQ}}\geq 9$. Robust solutions with $\Delta_{\cancel{PQ}} = 12$ have been identified, and we shall show now that with appropriate charge assignments it is possible to have models with $\Delta_{\cancel{PQ}} $ up to 18.

It is convenient to adopt a notation such that $(\psi_i\chi_i)^{\kappa_i}= (\psi_i^*\chi_i^*)^{\vert\kappa_i\vert}$ for $\kappa_i<0$, with the Lorentz, $G$ and $G_{\rm SM}$ indices contracted so that the pair $\psi_i\chi_i$ is a scalar operator, gauge singlet of $G$ and $G_{\rm SM}$, and $\kappa_i$ is an integer.

In this class of models, any anomalous $U(1)$ global symmetry corresponding to a PQ symmetry always acts as a multiple of the identity on the subsets $\{\psi_{i,\alpha},\chi_{i,\alpha}\}$ of the fields $\{\psi_{i},\chi_{i}\}$, with $\psi_{i,\alpha}$ corresponding to the irreducible $G_{\rm SM}$ representation $r_{i,\alpha}$, and similarly for $\chi_{i,\alpha}$. Therefore, given an operator of the schematic form $\prod_i (\psi_i\chi_i)$ it is always possible to construct a PQ violating one with the same dimension, by restricting to a single subset of fields on which the PQ symmetry generator acts nontrivially. 
On the other hand, according to the analysis of~\cite{Contino:2021ayn}, the PQ violating operators that generate a potential for the QCD axion are gauge invariant operators of the form $\prod_i (\psi_{i,\alpha}\chi_{i,\alpha})^{\kappa_i} $, possibly nonlocal products of local gauge invariant scalar operators. Therefore it is sufficient to characterise the set of operators of the form $\prod_i (\psi_i\chi_i)$.

An operator $\mathcal{O}_{s} = \prod_{i = 1}^{\nu} (\psi_i\chi_i)^{\kappa_i}$ is gauge invariant, \emph{i.e.} neutral under $U(1)$, if and only if
\begin{equation}\label{eq:pqv}
\sum_{i =1}^{\nu} \kappa_i (p_i + q_i)=0.
\end{equation}
At the level of a single insertion, the lowest dimensional operators generating a potential for the axion will then have a classical scaling dimension of
\begin{equation}\label{eq:mind}
\Delta_{s} =3 \cdot  {\rm Min}_{\{\kappa_i \}}\Bigg\{ \, \sum_{i = 1}^{\nu} |\kappa_i| \, \Bigg\},
\end{equation}
where the minimisation occurs among the set of integers $\{\kappa_i \}$ satisfying \eqref{eq:pqv}. In turn, one can determine the operator dimension for multiple insertions by performing the sum \eqref{eq:smi}.

\subsection{Models with $n_{f}=2$}

We focus now on models with $n_{f}=2$, whose classification along the lines previously discussed is summarised in Tab.~\ref{tab:nf2}. From now on we shall consider the case $d_2>d_1$, with $p_i+q_i \neq 0$, of interest for the solution of the QCD axion quality problem.

Regarding the dimension of PQ violating operators, Ref.~\cite{Contino:2021ayn} identified some universal PQ violating local operators which imply an upper bound on the dimension of the leading PQ violating operator $\Delta_{\cancel{PQ}}^{\rm max}$. The question of whether there are charge assignments with $\Delta_{\cancel{PQ}} = \Delta_{\cancel{PQ}}^{\rm max} $ was however left unsettled. Using our explicit solution we are able now to address this question, showing that in all the models  there are charge assignments with $\Delta_{\cancel{PQ}}=\Delta_{\cancel{PQ}}^{\rm max}$, thus providing a positive answer. In particular, for the case $n_f=2$, there are phenomenologically viable models with $\Delta_{\cancel{PQ}}=15$, that can therefore provide a particularly robust solution to the axion quality problem.

At the level of local operators, the condition \eqref{eq:pqv} together with equation \eqref{eq:gn2} gives:
\begin{equation}
\kappa_1 (6 d_2^2  k \ell) + \kappa_2 (-6 d_1d_2  k \ell) =0.
\end{equation}
Its unique nonzero solution for $k,\ell\neq 0$ is $[\kappa_1:\kappa_2] = [d_1:d_2] $, and corresponds to the PQ violating operators already identified in~\cite{Contino:2021ayn}. 
This shows that those operators are the lowest dimensional among \emph{local} operators. Their explicit form and classical dimension are
\begin{equation}\label{eq:PQviolating_nf2}
\mathcal{O}_s = (\psi_1\chi_1)^{\kappa_1} (\psi_2\chi_2)^{\kappa_2}, \qquad  \Delta_s = 3 (d_{1}+d_{2})/{\rm gcd}(d_{1},d_{2}) > 6.
\end{equation}

The question left to answer is whether it is possible to have a nonlocal operator $\mathcal{O}_{\rm PQ}$ of the form $(\psi_1\chi_1)^{\kappa_1} (\psi_2\chi_2)^{\kappa_2}$ which is the product of two or more local gauge invariant scalar operators and has a smaller effective dimension. We can always write this operator as the product of two gauge invariant scalar operators, $\mathcal{O}_{\rm PQ}=\mathcal{O}_{1}\mathcal{O}_{2}$, which are possibly nonlocal. \footnote{{\it{i.e.}} they can themselves be the result of a multiple insertion.} A net zero charge under $\U(1)$ is a necessary but not a sufficient condition for these operators to be gauge invariant, since specific choices of representations under $\GSM$ will impose additional requirements; however, these are model dependent and will not be considered in our analysis. Since $\kappa_1, \kappa_2$ have the same sign, the only possible suboperators which can be gauge singlets for every choice of group $G$ are $(\psi_i\chi_i)^{\lambda}$, $(\psi_1\chi_2)^{\lambda}, (\psi_2\chi_1)^{\lambda}$, or a product of them.\footnote{It is straightforward to check that additional factors of the form $(\psi_{i}^{\dagger}\psi_{j})$, $(\chi_{i}^{\dagger}\chi_{j})$ can be always neglected when interested in the lowest dimensional PQ violating operators.} The case $\mathcal{O}_{1}= (\psi_i\chi_i)^{\lambda}$ is excluded unless $k =0$ or $\ell =0$, which correspond to vectorlike assignments. Similarly, $\mathcal{O}_{1}= (\psi_i\chi_j)^{\lambda}$, with $i\neq j$ gives
\begin{equation}
\lambda (p_i+q_j) =0,
\end{equation}
which implies
\begin{equation}
d_2 \, k^2 \pm (d_2-d_1) k\ell=0,
\end{equation}
and has no solution for $k\neq0$, assuming $[k:\ell] \neq [\pm (d_2-d_1):d_2]$.  This also excludes the case $\mathcal{O}_{1}= (\psi_i\chi_i)^{\lambda_{1}} (\psi_j\chi_k)^{\lambda_2}$, since the operator $\mathcal{O}_{2}$ would be of the form we already excluded.
We are left with operators of the form $\mathcal{O}_{1} = (\psi_1\chi_2)^{\lambda_1}(\psi_2\chi_1)^{\lambda_2}$, and correspondingly $\mathcal{O}_{2} = (\psi_1\chi_1)^{\kappa_1-\lambda_1}(\psi_2\chi_2)^{\kappa_2-\lambda_1}(\psi_2\chi_1)^{\lambda_1-\lambda_2}$, with $[k_1:k_2]=[d_1:d_2]$. Their gauge invariance implies the condition:
\begin{equation}
\lambda_1 (p_1+q_2) + \lambda_2 (p_2+q_1)=0 \quad \implies \quad  \lambda_1 (k \, d_2+\ell (d_2-d_1)) + \lambda_2 (-k \, d_2+\ell (d_2-d_1)) =0,
\end{equation}
which is solved by $[\lambda_1:\lambda_2] = [k \, d_2-\ell (d_2-d_1): k \, d_2+\ell (d_2-d_1)]$. Given a choice of dimensions $d_1,d_2$, it is always possible to find a choice of $k,\ell$ such that the classical dimension of $\mathcal{O}_1$ considered as a local operator $\Delta(\mathcal{O}_{1})$ is arbitrarily large.
Indeed, for $k >\ell >0$,
\begin{equation}
\Delta(\mathcal{O}_{1}) = 3 \cdot {\rm Min}_{\{\lambda_i \}}\Big\{ \vert \lambda_1 \vert + \vert \lambda_2 \vert  \Big\} \geq 3\cdot \dfrac {2 k d_2}{2\ell (d_2-d_1)} \equiv \Delta(\mathcal{O}_{1})\Big\vert_{\rm min} ,
\end{equation}
since for $(\bar\lambda_1= k \, d_2-\ell (d_2-d_1) , \bar\lambda_2  =k \, d_2+\ell (d_2-d_1))$, we have that ${\rm gcd}(\bar\lambda_1,\bar\lambda_2)$ must be a divisor of $\bar\lambda_2 - \bar\lambda_1 = 2\ell (d_2-d_1)$. In the limit of large $k/\ell$ we can therefore increase the dimension of all operators with the same form as $\mathcal{O}_{1}$ arbitrarily, so that their insertions can be neglected.

On the other hand $\Delta(\mathcal{O}_{2})\geq 9$ if $\mathcal{O}_{2}$ is a local operator, so that even in the case in which $\Delta(\mathcal{O}_1)>9$ and it can be obtained as a nonlocal composite operator built as the product of local operators $\mathcal{O}_{2}$, its effective dimension is bounded by
\begin{equation}
\Delta(\mathcal{O}_{1})\Big\vert_{\rm eff} \geq \Delta(\mathcal{O}_{1})\Big\vert_{\rm min} - 4 \left(\dfrac{1}{9} \Delta(\mathcal{O}_{1})\Big\vert_{\rm min}  -1\right) = 4 + \dfrac{5}{9} \Delta(\mathcal{O}_{1})\Big\vert_{\rm min},
\end{equation}
which can again be made arbitrarily large by appropriately choosing $k/\ell$.

As a consequence, it is always possible to find charge assignments such that the leading PQ--violating effects are associated to local operators of the form~\eqref{eq:PQviolating_nf2}, with classical dimension $\Delta_{s}$. In particular, all the $n_f=2$ models classified in Tab.~\ref{tab:nf2} admit charge assignments with the maximal level of protection, $\Delta_{\cancel{PQ}} =\Delta_{\cancel{PQ}}^{\rm max} $. 

\begin{table}
\centering
\begin{tabular}{cl|cc|cc|c|c|c}
& $\GSM$  & $r_1$ & $r_2$ & $d_1$ & $d_2$ & & $\Delta^{\rm max}_{\cancel{PQ}}$ & $(k/\ell)_{min}$ \\[0.1cm]
\hline
& $\SU (3)_{\rm c}$ & $(\mathbf{3} \oplus m \, \mathbf{1})$ & $(\mathbf{3} \oplus m \, \mathbf{1})$ & $3+m$ & $3+m$ & $1\leq m \leq 23$ & 6 &  \\
& & $(\mathbf{3} \oplus m \, \mathbf{1})$ & $2\, (\mathbf{3} \oplus m \, \mathbf{1})$ & $3+m$ & $2(3+m)$ & $1\leq m \leq 8$ & 9 &  \\
& & $(\mathbf{3} \oplus m \, \mathbf{1})$ & $3\, (\mathbf{3} \oplus m \, \mathbf{1})$ & $3+m$ & $3(3+m)$ & $1\leq m \leq 3$ & 12 & 14/9 \\
  & & $2\, (\mathbf{3} \oplus m \, \mathbf{1})$ & $2\, (\mathbf{3} \oplus m \, \mathbf{1})$ & $2(3+m)$ & $2(3+m)$ & $1\leq m \leq 3$ & 6 & \\
  & & $(\mathbf{3} \oplus \mathbf{1})$ & $4\, (\mathbf{3} \oplus \mathbf{1})$ & $4$ & $16$ & & 15 & 5/2 \\
  & & $2\, (\mathbf{3} \oplus \mathbf{1})$ & $3\, (\mathbf{3} \oplus \mathbf{1})$ & $8$ & $12$ & & 15 & 10/9  \\
\hline
\hline
& $\rm SU(5)_{GUT}$  & $(\mathbf{5} \oplus m \, \mathbf{1})$ & $(\mathbf{5} \oplus m \, \mathbf{1})$ & $5+m$ & $5+m$ & $1\leq m \leq 21$ & 6 & \\
  & & $(\mathbf{5} \oplus m \, \mathbf{1})$ & $2\, (\mathbf{5} \oplus m \, \mathbf{1})$ & $3+m$ & $2(3+m)$ & $1\leq m \leq 6$ & 9 &  \\
  & & $(\mathbf{5} \oplus \mathbf{1})$ & $3\, (\mathbf{5} \oplus \mathbf{1})$ & $6$ & $18$ & & 12 & 14/9 \\
  & & $2\, (\mathbf{5} \oplus \mathbf{1})$ & $2\, (\mathbf{5} \oplus \mathbf{1})$ & $12$ & $12$ & & 6 & \\
& & $\mathbf{5}$ & $(\mathbf{10} \oplus 5 \, \mathbf{1})$ & $5$ & $15$ & & 12 & 14/9 \\
\end{tabular}
\caption{\it Classification of composite axion models with $n_{f}=2$, reproduced from~\cite{Contino:2021ayn}. The multiplicity of SM singlets $m$ is an integer in the specified range. $\Delta_{\cancel{PQ}}^{\rm max}$ denotes the dimension of the leading single insertion PQ violating operators, while in the last column we list a lower bound on $k/\ell$, for $k>\ell>0$, that ensures that the effect of multiple insertions is negligible. An empty slot means that single insertions are always the leading effect for that choice of representation. The corresponding charges can be obtained from eq.~\eqref{eq:gn2}.}
\label{tab:nf2}
\end{table}

\subsection{Models with $n_{f}=3$ and beyond}

For considerations related to PQ violating operators it is convenient to use the parametrisation of eq.~\eqref{eq:sol_general} for the $U(1)$ charges, in the case where $n_{f} >2$. For this reason, we shall first provide a unified treatment, and later specialise to the particular case $n_{f}=3$. In this section we set again $\nu \equiv n_f$ for notational convenience.
Following this approach, eq. \eqref{eq:pqv} can be written as 
\begin{equation}\label{eq:pqntr}
\sum_{i =1}^{\nu-1} (\ell_i + m_i) (\kappa_i d_{\nu}- d_i \kappa_{\nu}) = 0,
\end{equation}
supplemented by the constraint 
\begin{equation}\label{eq:ctr}
\sum_{i =1}^{\nu -1} (\ell_i + m_i) D_{i \nu} = 0.
\end{equation}
We remind the reader that the parameters describing the solutions \eqref{eq:sol_general} are given by the set $\{\ell_i, m_i \}_{i= 1,...,\nu -1}$ plus the additional variable $\ell_{\nu}$, which does not appear in the two equations above. The solutions with $\ell_i = - m_i$ correspond to vectorlike charge assignments, and we shall henceforth assume $\ell_i + m_i \neq 0$.

Substituting the constraint \eqref{eq:ctr} into \eqref{eq:pqntr}, one is left with
\begin{equation}\label{eq:lmq}
\sum_{i =1}^{\nu-2} (\ell_i + m_i) \big[D_{\nu-1 \, \nu} (\kappa_i d_{\nu}- d_i \kappa_{\nu}) -D_{i \, \nu} (\kappa_{\nu-1} d_{\nu}-d_{\nu-1} \kappa_{\nu})  \big] = 0,
\end{equation}
an expression only in terms of the independent parameters $\{\ell_i, m_i \}_{i= 1,...,\nu -2}$. The case $\nu=n_f = 3$ gives a linear Diophantine equation in the three integer variables $\kappa_i$ and this can be easily solved once a choice of representations has been specified. It is given by
\begin{equation}\label{eq:n3k}
\kappa_1 D_{23}+\kappa_2 D_{31}+\kappa_3 D_{12}=0.
\end{equation}
This shows in particular that at the level of single insertion, the existence of PQ violating operators and their leading dimension is determined only by the choice of $G_{\rm SM}$ representations and does not depend on the specific assignments of U(1) charges. This feature is true also for the $n_f=2$ models of the previous section, and was also noticed \emph{a posteriori} in the numerical scan of PQ violating operators performed in~\cite{Contino:2021ayn} for selected $n_f=3$ models.

In general, defining $E = {\rm gcd}(D_{23},D_{31})$, and given an explicit solution $(x_0,y_0)$ to the equation
\begin{equation}
D_{23} \,x + D_{31}\, y = E,
\end{equation} 
which can always be found through Euclid's algorithm, the general solution to eq.~\eqref{eq:n3k} is:
\begin{equation}
[\kappa_1:\kappa_2:\kappa_3] = 
[-D_{12} (x_0 h_1 + D_{31} h_2)
: -D_{12} (y_0 h_1 - D_{23} h_2)
: E h_1],
\end{equation}
with $h_1,h_2\in \mathbb{Z}$ arbitrary integers. Moreover, an obvious solution to eq.~\eqref{eq:n3k} is $(\kappa_1,\kappa_2,\kappa_3) = (d_1,d_2,d_3)$, which allows one to extract the upper bound
\begin{equation}
\Delta_s \leq 3 \frac{d_1+d_2+d_3}{{\rm{gcd}}(d_1,d_2,d_3)}.
\end{equation}
This is however hardly saturated in practice, as one can usually find solutions with smaller $\kappa_i$'s.

We are ready now to apply this general analysis to the case of the composite axion models with $n_f=3$ and irreducible $\GSM $ representation classified in Ref.~\cite{Contino:2021ayn}, as summarised in Tab.~\ref{tab:nf3}. As an example, consider the model with $\GSM = \SU(5)_{\rm{GUT}}$ and $(r_1,r_2,r_3)=(\mathbf{1},\mathbf{\bar{5}} ,\mathbf{10})$ or $(r_1,r_2,r_3)=(\mathbf{1},\mathbf{5} ,\mathbf{10})$. There,  eq. \eqref{eq:n3k} reads
\begin{equation}
5 \kappa_1 -3 \kappa_2 + \kappa_3 = 0.
\end{equation}
Its general solution is $[\kappa_1:\kappa_2:\kappa_3] = [h_1 +3 h_2: 2 h_1 +5 h_2 : h_1] $.
The solutions which minimise the expression in \eqref{eq:mind} are obtained for $(h_1,h_2)$ equal to $(+1,0)$, $(-2,+1)$ and $(3,-1)$, corresponding to $(\kappa_1,\kappa_2,\kappa_3)= (+1,+2,+1)$, $(\kappa_1,\kappa_2,\kappa_3)= (+1,+1,-2)$ and $(\kappa_1,\kappa_2,\kappa_3)= (0,+1,+3)$ respectively, and result in $\Delta_s=12$. These operators coincide with the generic PQ violating operators identified in Ref.~\cite{Contino:2021ayn}, present irrespectively of the choice of $U(1)$ charges. The absence of additional local operators for arbitrary values of the charges was a mysterious output of the numerical scan that we now understand from an algebraic perspective. It is straightforward to repeat the same analysis for the other choices of representations.

\begin{table}
\centering
\begin{tabular}{cl|ccc|ccc|c}
& $\GSM$  & $r_1$ & $r_2$ & $r_3$ & $D_{12}$ & $D_{23}$ & $D_{31}$ & $\Delta^{\rm max}_{\cancel{PQ}}$ \\[0.1cm]
\hline  
  & $\SU(3)_{c}$ & $\mathbf{1}$ & $\mathbf{3}$ &  $\mathbf{6}$ & 1 & 9& -5 &  $12$  \\
&  &  $\mathbf{8}$ & $\mathbf{3}$ & $\mathbf{6}$ & -10 & 9& -4 & $15$   \\
  &  & $\mathbf{1}$ & $\mathbf{3}$ &  $\mathbf{8}$ & 1 & 10& -6 &  $15$  \\
&  & $\mathbf{1}$ & $\mathbf{6}$ &  $\mathbf{8}$ & 5 & -4& -6 &   $12$  \\
\hline
\hline
& $\rm SU(5)_{GUT}$ & $\mathbf{1}$ & $\mathbf{\bar{5}}$ &  $\mathbf{10}$ & 1 & 5& -3 &  $12$ \\
  & & $\mathbf{1}$ & $\mathbf{\bar 5}$ &  $\mathbf{15}$ & 1 & 20& -7 &  $15$  \\
  & & $\mathbf{1}$ & $\mathbf{10}$ &  $\mathbf{15}$ & 3 & 25& -7 &  $18$  \\
\end{tabular}
\caption{\it Classification of composite axion models with $n_{f}=3$ and irreducible representations, reproduced from~\cite{Contino:2021ayn}. The dimension $d_i$ of the representation $r_i$ can be read from the symbol corresponding to the irreducible representation. We report the values of the coefficients $D_{ij}$ defined in eq.~\eqref{eq:dij}. The last column reports the dimension of the leading single insertion PQ violating operator, as confirmed from our analysis. For the charge assignments specified by parameters satisfying eq.~\eqref{eq:clm2}, multiple insertions are negligible and $\Delta_{\cancel{PQ}} =\Delta_{\cancel{PQ}}^{\rm max} $.}
\label{tab:nf3}
\end{table}

For $\nu=n_f >3$, it is always possible to find an assignment of the $\{ \ell_i, m_i \} $ such that, for any choice of the $\kappa_i$ within a finite set, eq. \eqref{eq:lmq} can only be satisfied if it vanishes term by term, {\it{i.e.}} 
\begin{equation}\label{eq:n3si}
J_i (\kappa) \equiv D_{\nu-1 \, \nu} (\kappa_i d_{\nu}- d_i \kappa_{\nu}) -D_{i \, \nu} (\kappa_{\nu-1} d_{\nu}-d_{\nu-1} \kappa_{\nu})  = 0, \quad \quad i = 1,..., \nu-2.
\end{equation}
To do so, one starts by fixing a desired level of protection $\Delta^*$ that is required at the level of single insertions, and correspondingly chooses the $\kappa_i$ to be lying within the set 
\begin{equation}\label{eq:kappastar}
\mathcal{K} (\Delta^*)= \{\, \kappa_i \in \mathbb{Z}, \,\, \rm{with} \,\,  3 \sum |\kappa_i | \leq \Delta^*   \,\}.
\end{equation}
A suitable choice of the $\{\ell_i,m_i \}$ can then be constructed as follows. Starting from arbitrary values of $(\ell_1, m_1)$, the elements $\{ \ell_i, m_i \}_{i=1...\nu-2} $ should be chosen to satisfy 
\begin{equation}\label{eq:clm1}
|\ell_j + m_j |>  {\rm max}_{\{\kappa_i \in \mathcal{K} (\Delta^*),\,  J_j(\kappa) \neq 0 \}}\Bigg\{ \,  
\Bigg| \frac{ \sum_{i =1}^{j-1} (\ell_i + m_i) J_i(\kappa)}{J_j(\kappa) } \, \Bigg| \Bigg\},
\end{equation}
which is well defined for any value of $\Delta^*$ since the right-hand side is a bounded quantity (the denominator is a nonvanishing integer) and the set $\mathcal{K}$ is finite. Working in reverse, it is now easy to see that all the $J_i$'s in \eqref{eq:n3si} must vanish if eq. \eqref{eq:lmq} is to be satisfied, starting from $i = \nu -2$ and proceeding in descending order up to $i =1$.

With a few algebraic manipulations, the equations in \eqref{eq:n3si} can be turned into the system
\begin{equation}\label{eq:general_system_kappa}
\begin{cases}
\displaystyle k_1 D_{\nu-1 \, \nu}+ k_{\nu-1} D_{\nu \, 1}+  k_{\nu} D_{1 \, \nu-1}  = 0  \\
\displaystyle \quad \quad \quad \quad \quad \quad \quad \vdots\\
\displaystyle k_i \,D_{\nu-1 \, \nu}+ k_{\nu-1} D_{\nu \, i}+  k_{\nu} \,D_{i \, \nu-1}  = 0  \\
\displaystyle \quad \quad \quad \quad \quad \quad \quad \, \vdots\\
\displaystyle k_{\nu-2} D_{\nu-1 \, \nu}+ k_{\nu-1} D_{\nu \, \nu-2}+  k_{\nu} D_{\nu-2 \, \nu-1}  = 0.  \\
\end{cases}
\end{equation}
The first equation coincides with \eqref{eq:n3k} after a relabeling, and is therefore solved by
\begin{equation}\label{eq:gs1}
[\kappa_1:\kappa_{\nu-1}:\kappa_{\nu}] = 
[-D_{1\,\nu-1} (x_0 h_1 + D_{\nu \, 1} h_2)
: -D_{1\, \nu-1} (y_0 h_1 - D_{\nu \, \nu-1} h_2)
: E_{\nu} h_1],
\end{equation}
where  $E_{\nu} = {\rm gcd}(D_{\nu \, \nu-1},D_{\nu 1})$ and, as before, $(x_0,y_0)$ is a particular solution of
\begin{equation}
D_{\nu \, \nu-1}\, x + D_{\nu 1} \,y = E_{\nu}.
\end{equation}
The remaining $\nu-3$ variables can be extracted from the corresponding equations as
\begin{gather}\label{eq:gs2}
[\kappa_i:\kappa_{\nu-1}:\kappa_{\nu}] = \\
[D_{\nu\, i}  D_{1\, \nu-1} (y_0 h_1 - D_{\nu \, \nu-1} h_2)-D_{i \, \nu-1} E_{\nu} h_1
: -D_{\nu-1\, \nu} D_{1\, \nu-1} (y_0 h_1 - D_{\nu \, \nu-1} h_2)
: D_{\nu-1\, \nu} E_{\nu} h_1].
\end{gather}
Analogously to the case $n_f =2$, one can then define a maximum level of protection 
\begin{equation}
\Delta_{\cancel{PQ}}^{\rm max}  =3 \cdot  {\rm Min}_{\{\kappa_i \}}\Bigg\{ \, \sum_{i = 1}^{\nu} |\kappa_i| \, \Bigg\},
\end{equation}
where the minimization is restricted to the $\{\kappa_i\}$ taking values in \eqref{eq:gs1} and \eqref{eq:gs2}. For $n_{f} = 3$, this coincides with the $ \Delta_{\cancel{PQ}}^{\rm max} $ of \cite{Contino:2021ayn}, corresponding to the operators discussed in the previous paragraph. Setting $\Delta^* = \Delta_{\cancel{PQ}}^{\rm max} $ in eq.\eqref{eq:kappastar}, we have then shown that there exist charge assignments for which all single insertions have a dimension $\Delta_s \geq \Delta_{\cancel{PQ}}^{\rm max}$.

A complete analysis of the effects due to multiple insertions would probably require a numerical, brute-force approach to implement $\GSM$ gauge invariance, as performed in \cite{Contino:2021ayn} for a specific choice of representations. However, we shall show that there always exist values of the $\{\ell_i,m_i \}$ (and hence of the $\U(1)$ charges) for which the effect of multiple insertions is subdominant with respect to that of single insertions, so that $\Delta_{\cancel{PQ}} =\Delta_{\cancel{PQ}}^{\rm max} $ can be obtained even for the models with general $n_{f}$. In particular, this applies to the $n_{f} =3 $ models listed in Table \ref{tab:nf3}. The statement can be shown by proving that for such values of the charges it is not possible to write down any local, gauge invariant operator with a dimension smaller then those of the form $\mathcal{O}_{s} = \prod_{i = 1}^{\nu} (\psi_i\chi_i)^{\kappa_i}$. It then follows from \eqref{eq:smi} that any multiple insertion will have an effective dimension  of $\Delta_{\rm{eff}} \geq N (\Delta_{\cancel{PQ}}^{\rm max} -4)+ 4 \geq \Delta_{\cancel{PQ}}^{\rm max} $.

The fermion content of a generic operator can be written as $\mathcal{O}_{\varepsilon} = \prod_{i = 1}^{\nu} \psi_i^{\kappa_i+ \varepsilon_i } \chi_i^{\kappa_i }$, where gauge invariance under $G$ for arbitrary gauge group $G$ imposes the constraint $\sum_{i=1}^{\nu} \varepsilon_i =0$, and we do not impose invariance under $\GSM$. Generalising \eqref{eq:pqv}, the requirement of neutrality under $\U(1)$ can be stated as
\begin{equation}\label{eq:ng}
\sum_{i =1}^{\nu} \kappa_i (p_i +q_i) + \varepsilon_i p_i =0.
\end{equation}
Thus, the lowest dimension they can attain is given by
\begin{equation}
\Delta_{\varepsilon} =\frac{3}{2} \cdot  {\rm Min}_{\{\kappa_i, \varepsilon_i \}}\Bigg\{ \, \sum_{i = 1}^{\nu} |\kappa_i| + \sum_{i = 1}^{\nu} |\kappa_i+ \varepsilon_i| \, \Bigg\},
\end{equation}
where the minimum is evaluated over all values of $\{\kappa_i, \varepsilon_i \}$ satisfying \eqref{eq:ng}. Of course $\Delta_{\varepsilon} \leq \Delta_{\cancel{PQ}}^{\rm max} $, corresponding to the case where $\varepsilon_i =0$ for all $i$. Using the fact that $\sum_{i=1}^{\nu} \varepsilon_i =0$, eq. \eqref{eq:ng} can be turned into (assuming $A_1 \neq 0$)
\begin{equation}\label{eq:kep}
\sum_{i = 1}^{\nu-2} J_i(\kappa) (\ell_i + m_i) + d_{\nu}\sum_{i=1}^{\nu} \varepsilon_i \ell_i = 0;
\end{equation}
the first term can be recognised as the one corresponding to single insertions. We now wish to prove the existence of suitable charge assignments, for which no solutions to eq. \eqref{eq:kep} can realize $\Delta_{\varepsilon} < \Delta_{\cancel{PQ}}^{\rm max} $. The strategy is to choose values of the $\{\ell_i,m_i\}$ such that eq. \eqref{eq:kep} must be satisfied term by term (with a similar procedure to the one used for single insertions):
\begin{equation}\label{eq:fs}
J_i(\kappa)= 0 \quad \quad \quad  \quad \quad \quad \sum_{i=1}^{\nu} \varepsilon_i \ell_i = 0,
\end{equation}
and such that there are no solutions to \eqref{eq:fs} with $\Delta_{\varepsilon} < \Delta_{\cancel{PQ}}^{\rm max} $. They are defined as those belonging to the set 
\begin{equation}
\mathcal{K}_{\varepsilon} (\Delta^{\rm max}_{\cancel{PQ}})= \Big\{\, \kappa_i,\varepsilon_i \in \mathbb{Z}, \,\, \rm{with} \,\,  \frac{3}{2} \Big( \sum |\kappa_i |+\sum |\kappa_i +\varepsilon_i| \Big) < \Delta^{\rm max}_{\cancel{PQ}}  \,\Big\}.
\end{equation}
To ensure the latter condition is satisfied, one starts by choosing the $\{\ell_i \}$ in such a way that the elements of $\mathcal{K}_{\varepsilon}( \Delta_{\cancel{PQ}}^{\rm max} )$ can never solve the independent equations in \eqref{eq:fs} simultaneously. This can always be done by taking the $\ell_i$ to be large enough, for instance $\ell_{j+1} > | \sum_{i=1}^{j} \varepsilon_i \ell_i|$ for any $\varepsilon_i \in \mathcal{K}_{\varepsilon}(\Delta^{\rm max}_{\cancel{PQ}})$.\footnote{The set $\mathcal{K}_{\varepsilon} (\Delta^{\rm max}_{\cancel{PQ}})$ is finite, so this prescription is unambiguous. } This implies eq. \eqref{eq:fs} can be satisfied only if the $\varepsilon_i $ vanish one by one starting from $i=\nu$, a solution which does not belong to $\mathcal{K}_{\varepsilon}( \Delta_{\cancel{PQ}}^{\rm max} )$. On the other hand, the $\{m_i\}_{i =1...\nu-2}$ can be chosen in succession so that they satisfy\footnote{At this point, $m_{\nu-1}$ will be univocally determined by the constraint \eqref{eq:ctr}.}
\begin{equation}\label{eq:clm2}
|\ell_j + m_j |>  {\rm max}_{\{\kappa_i, \varepsilon_i \in \mathcal{K}_{\varepsilon} (\Delta^{\rm max}_{\cancel{PQ}}),\,  J_j(\kappa) \neq 0 \}}\Bigg\{ \,  
\Bigg| \frac{ \sum_{i =1}^{j-1} (\ell_i + m_i) J_i(\kappa) }{J_j(\kappa) } \, \Bigg|
+\Bigg| \frac{d_{\nu} \sum_{i=1}^{\nu} \varepsilon_i \ell_i  }{J_j(\kappa) } \, \Bigg| \Bigg\},
\end{equation}
in such a way that, in analogy to what was argued previously for eq.~\eqref{eq:clm1}, any solution of \eqref{eq:kep} with $\kappa_i, \varepsilon_i \in \mathcal{K}_{\varepsilon} (\Delta^{\rm max}_{\cancel{PQ}})$ requires all terms $J_i(\kappa)$ to vanish individually, implying \eqref{eq:fs}.
By construction, the solutions of \eqref{eq:fs} cannot belong to $\mathcal{K}_{\varepsilon} (\Delta^{\rm max}_{\cancel{PQ}})$, and one concludes that there cannot be gauge invariant, local operators with a dimension smaller than $\Delta_{\cancel{PQ}}^{\rm max}$ (corresponding to the already identified single insertion).\footnote{Except the trivial operators solely built out of $(\psi_{i}^{\dagger}\psi_{j})$, $(\chi_{i}^{\dagger}\chi_{j})$ factors, that do not affect the axion potential.} Notice that eq. \eqref{eq:clm2} automatically implies \eqref{eq:clm1}, so that single and multiple insertions can be accounted for simultaneously, and the charge assignments now specified indeed have $\Delta_{\cancel{PQ}} =\Delta_{\cancel{PQ}}^{\rm max} $. In particular, all the models of Tab.~\ref{tab:nf3} admit charge assignments with the maximum level of protection.

\section*{Acknowledgments}
It is a pleasure to thank Roberto Contino for collaboration on related subjects and for comments. A.P. thanks Frederik Denef for comments on rational varieties, and acknowledges support from the Simons Foundation Award No.~658906. F.R. acknowledges support from the Dalitz Graduate Scholarship, jointly established by the Oxford University Department of Physics and Wadham College. 

\appendix

\section{Points at infinity for $n_f=3$}\label{app:A}

We collect a few explicit examples concerning the additional family of solutions that may arise from points at infinity for $n_f=3$, showing how all the three possibilities listed at the end of Section \ref{ssc:n3} can be realised in practice. These models do not have any pretense of being realistic or phenomenologically relevant, but only serve as a proof of principle for the existence of such solutions. For this reason, we can take the would-be-SM gauge group $G_{\rm SM}$ as any semisimple Lie group.

In the notation of Section \ref{ssc:n3}, points at infinity correspond to the solutions of eq. (\ref{eq:W0}), which can also be parametrised as in (\ref{eq:pt}). The different representations can always be arranged in such a way that $b \cdot c >0$, leading to three qualitatively distinct cases. In each of these, we write down a choice of representations that would realise that possibility, and the corresponding form of equation (\ref{eq:W0}).

\paragraph{Case 1:}$a \cdot b \geq 0 $, no solution:
\begin{equation}
G_{{\rm{SM}}} = \SU(3), \quad \quad r_1= \mathbf{3}, \quad \quad r_2= \mathbf{6}, \quad \quad r_3= \mathbf{1},
\end{equation}
\begin{equation}
8 X^2+10 Y^2+15 Z^2=0 .
\end{equation}
Being a sum of positive terms, this equation does not admit any nontrivial integer solutions.

\paragraph{Case 2:}$a \cdot b < 0$, infinite solutions:
\begin{equation}
G_{{\rm{SM}}} = \SU(4), \quad \quad r_1= \mathbf{4}, \quad \quad r_2= \mathbf{1}, \quad \quad r_3= \mathbf{6},
\end{equation}
\begin{equation}
3X^2-4Y^2-12 Z^2=0.
\end{equation}
One can check by inspection that the point $(X_0,Y_0,Z_0)=(2,0,1)$ belongs to the conic. With the methods employed in the text, the general solution can be written down as
\begin{equation}
X = \frac{n}{\mu}(8 k^2 + 6  \ell^2), \quad \quad Y = \frac{n}{\mu} 12 k \,\ell,\quad \quad Z = \frac{n}{\mu}(4k^2-3\ell^2), \quad \quad n,k,\ell \in \mathbb{Z},
\end{equation}
with
\begin{equation}
\mu = {\rm{gcd}} \big(8 k^2 + 6  \ell^2,12 k\, \ell,4k^2-3\ell^2 \big).
\end{equation}

\paragraph{Case 3:}$a \cdot b < 0$, no solution
\begin{equation}
G_{{\rm{SM}}} = \SU(5), \quad \quad r_1= \mathbf{10}, \quad \quad r_2= \mathbf{5}, \quad \quad r_3= \mathbf{24},
\end{equation}
\begin{equation}
525 X^2 - 2366 Y^2 - 2028 Z^2 = 0.
\end{equation}
One can check numerically that there are no integer solutions in the domain defined by (\ref{eq:int}), so that the only one is the trivial one.


\phantomsection
\addcontentsline{toc}{section}{References}
\bibliographystyle{utphys}
{\linespread{1.075}
\bibliography{biblio}

\providecommand{\href}[2]{#2}\begingroup\raggedright\begin{thebibliography}{10}

\bibitem{Weinberg:1996kr}
S.~Weinberg, {\em {The quantum theory of fields. Vol. 2: Modern applications}}.
\newblock Cambridge University Press, 8, 2013.

\bibitem{Banks:2010zn}
T.~Banks and N.~Seiberg, ``{Symmetries and Strings in Field Theory and
  Gravity},'' \href{http://dx.doi.org/10.1103/PhysRevD.83.084019}{{\em Phys.
  Rev. D} {\bfseries 83} (2011) 084019},
  \href{http://arxiv.org/abs/1011.5120}{{\ttfamily arXiv:1011.5120 [hep-th]}}.

\bibitem{Harlow:2018tng}
D.~Harlow and H.~Ooguri, ``{Symmetries in quantum field theory and quantum
  gravity},'' \href{http://dx.doi.org/10.1007/s00220-021-04040-y}{{\em Commun.
  Math. Phys.} {\bfseries 383} no.~3, (2021) 1669--1804},
  \href{http://arxiv.org/abs/1810.05338}{{\ttfamily arXiv:1810.05338
  [hep-th]}}.

\bibitem{Costa:2019zzy}
D.~B. Costa, B.~A. Dobrescu, and P.~J. Fox, ``{General Solution to the U(1)
  Anomaly Equations},''
  \href{http://dx.doi.org/10.1103/PhysRevLett.123.151601}{{\em Phys. Rev.
  Lett.} {\bfseries 123} no.~15, (2019) 151601},
  \href{http://arxiv.org/abs/1905.13729}{{\ttfamily arXiv:1905.13729
  [hep-th]}}.

\bibitem{Allanach:2018vjg}
B.~C. Allanach, J.~Davighi, and S.~Melville, ``{An Anomaly-free Atlas: charting
  the space of flavour-dependent gauged $U(1)$ extensions of the Standard
  Model},'' \href{http://dx.doi.org/10.1007/JHEP02(2019)082}{{\em JHEP}
  {\bfseries 02} (2019) 082}, \href{http://arxiv.org/abs/1812.04602}{{\ttfamily
  arXiv:1812.04602 [hep-ph]}}. [Erratum: JHEP 08, 064 (2019)].

\bibitem{Allanach:2019gwp}
B.~C. Allanach, B.~Gripaios, and J.~Tooby-Smith, ``{Geometric General Solution
  to the $U(1)$ Anomaly Equations},''
  \href{http://dx.doi.org/10.1007/JHEP05(2020)065}{{\em JHEP} {\bfseries 05}
  (2020) 065}, \href{http://arxiv.org/abs/1912.04804}{{\ttfamily
  arXiv:1912.04804 [hep-th]}}.

\bibitem{Costa:2020dph}
D.~B. Costa, B.~A. Dobrescu, and P.~J. Fox, ``{Chiral Abelian gauge theories
  with few fermions},''
  \href{http://dx.doi.org/10.1103/PhysRevD.101.095032}{{\em Phys. Rev. D}
  {\bfseries 101} no.~9, (2020) 095032},
  \href{http://arxiv.org/abs/2001.11991}{{\ttfamily arXiv:2001.11991
  [hep-ph]}}.

\bibitem{Allanach:2019uuu}
B.~C. Allanach, B.~Gripaios, and J.~Tooby-Smith, ``{Solving local anomaly
  equations in gauge-rank extensions of the Standard Model},''
  \href{http://dx.doi.org/10.1103/PhysRevD.101.075015}{{\em Phys. Rev. D}
  {\bfseries 101} no.~7, (2020) 075015},
  \href{http://arxiv.org/abs/1912.10022}{{\ttfamily arXiv:1912.10022
  [hep-th]}}.

\bibitem{Allanach:2020zna}
B.~C. Allanach, B.~Gripaios, and J.~Tooby-Smith, ``{Anomaly cancellation with
  an extra gauge boson},''
  \href{http://dx.doi.org/10.1103/PhysRevLett.125.161601}{{\em Phys. Rev.
  Lett.} {\bfseries 125} no.~16, (2020) 161601},
  \href{http://arxiv.org/abs/2006.03588}{{\ttfamily arXiv:2006.03588
  [hep-th]}}.

\bibitem{Costa:2020krs}
D.~B. Costa, ``{Anomaly-free $U(1)^m$ extensions of the Standard Model},''
  \href{http://dx.doi.org/10.1103/PhysRevD.102.115006}{{\em Phys. Rev. D}
  {\bfseries 102} no.~11, (2020) 115006},
  \href{http://arxiv.org/abs/2007.08733}{{\ttfamily arXiv:2007.08733
  [hep-ph]}}.

\bibitem{Dobrescu:2020evn}
B.~A. Dobrescu and P.~J. Fox, ``{Diophantine equations with sum of cubes and
  cube of sum},'' \href{http://dx.doi.org/10.4310/CNTP.2022.v16.n2.a4}{{\em
  Commun. Num. Theor. Phys.} {\bfseries 16} no.~2, (2022) 401--434},
  \href{http://arxiv.org/abs/2012.04139}{{\ttfamily arXiv:2012.04139
  [math.NT]}}.

\bibitem{Lohitsiri:2019fuu}
N.~Lohitsiri and D.~Tong, ``{Hypercharge Quantisation and Fermat's Last
  Theorem},'' \href{http://dx.doi.org/10.21468/SciPostPhys.8.1.009}{{\em
  SciPost Phys.} {\bfseries 8} no.~1, (2020) 009},
  \href{http://arxiv.org/abs/1907.00514}{{\ttfamily arXiv:1907.00514
  [hep-th]}}.

\bibitem{Kaplan:1983fs}
D.~B. Kaplan and H.~Georgi, ``{SU(2) x U(1) Breaking by Vacuum Misalignment},''
  \href{http://dx.doi.org/10.1016/0370-2693(84)91177-8}{{\em Phys. Lett. B}
  {\bfseries 136} (1984) 183--186}.

\bibitem{Kaplan:1983sm}
D.~B. Kaplan, H.~Georgi, and S.~Dimopoulos, ``{Composite Higgs Scalars},''
  \href{http://dx.doi.org/10.1016/0370-2693(84)91178-X}{{\em Phys. Lett. B}
  {\bfseries 136} (1984) 187--190}.

\bibitem{Dugan:1984hq}
M.~J. Dugan, H.~Georgi, and D.~B. Kaplan, ``{Anatomy of a Composite Higgs
  Model},'' \href{http://dx.doi.org/10.1016/0550-3213(85)90221-4}{{\em Nucl.
  Phys. B} {\bfseries 254} (1985) 299--326}.

\bibitem{Contino:2010rs}
R.~Contino, \href{http://dx.doi.org/10.1142/9789814327183_0005}{``{The Higgs as
  a Composite Nambu-Goldstone Boson},''} in {\em {Theoretical Advanced Study
  Institute in Elementary Particle Physics}: {Physics of the Large and the
  Small}}, pp.~235--306.
\newblock 2011.
\newblock \href{http://arxiv.org/abs/1005.4269}{{\ttfamily arXiv:1005.4269
  [hep-ph]}}.

\bibitem{Panico:2015jxa}
G.~Panico and A.~Wulzer,
  \href{http://dx.doi.org/10.1007/978-3-319-22617-0}{{\em {The Composite
  Nambu-Goldstone Higgs}}}, vol.~913.
\newblock Springer, 2016.
\newblock \href{http://arxiv.org/abs/1506.01961}{{\ttfamily arXiv:1506.01961
  [hep-ph]}}.

\bibitem{Kim:1984pt}
J.~E. Kim, ``{A COMPOSITE INVISIBLE AXION},''
  \href{http://dx.doi.org/10.1103/PhysRevD.31.1733}{{\em Phys. Rev. D}
  {\bfseries 31} (1985) 1733}.

\bibitem{Kaplan:1985dv}
D.~B. Kaplan, ``{Opening the Axion Window},''
  \href{http://dx.doi.org/10.1016/0550-3213(85)90319-0}{{\em Nucl. Phys. B}
  {\bfseries 260} (1985) 215--226}.

\bibitem{Choi:1985cb}
K.~Choi and J.~E. Kim, ``{DYNAMICAL AXION},''
  \href{http://dx.doi.org/10.1103/PhysRevD.32.1828}{{\em Phys. Rev. D}
  {\bfseries 32} (1985) 1828}.

\bibitem{Randall:1992ut}
L.~Randall, ``{Composite axion models and Planck scale physics},''
  \href{http://dx.doi.org/10.1016/0370-2693(92)91928-3}{{\em Phys. Lett. B}
  {\bfseries 284} (1992) 77--80}.

\bibitem{Dobrescu:1996jp}
B.~A. Dobrescu, ``{The Strong CP problem versus Planck scale physics},''
  \href{http://dx.doi.org/10.1103/PhysRevD.55.5826}{{\em Phys. Rev. D}
  {\bfseries 55} (1997) 5826--5833},
  \href{http://arxiv.org/abs/hep-ph/9609221}{{\ttfamily arXiv:hep-ph/9609221}}.

\bibitem{Redi:2016esr}
M.~Redi and R.~Sato, ``{Composite Accidental Axions},''
  \href{http://dx.doi.org/10.1007/JHEP05(2016)104}{{\em JHEP} {\bfseries 05}
  (2016) 104}, \href{http://arxiv.org/abs/1602.05427}{{\ttfamily
  arXiv:1602.05427 [hep-ph]}}.

\bibitem{Fukuda:2017ylt}
H.~Fukuda, M.~Ibe, M.~Suzuki, and T.~T. Yanagida, ``{A ''gauged'' $U(1)$
  Peccei\textendash{}Quinn symmetry},''
  \href{http://dx.doi.org/10.1016/j.physletb.2017.05.071}{{\em Phys. Lett. B}
  {\bfseries 771} (2017) 327--331},
  \href{http://arxiv.org/abs/1703.01112}{{\ttfamily arXiv:1703.01112
  [hep-ph]}}.

\bibitem{Lillard:2017cwx}
B.~Lillard and T.~M.~P. Tait, ``{A Composite Axion from a Supersymmetric
  Product Group},'' \href{http://dx.doi.org/10.1007/JHEP11(2017)005}{{\em JHEP}
  {\bfseries 11} (2017) 005}, \href{http://arxiv.org/abs/1707.04261}{{\ttfamily
  arXiv:1707.04261 [hep-ph]}}.

\bibitem{Lillard:2018fdt}
B.~Lillard and T.~M.~P. Tait, ``{A High Quality Composite Axion},''
  \href{http://dx.doi.org/10.1007/JHEP11(2018)199}{{\em JHEP} {\bfseries 11}
  (2018) 199}, \href{http://arxiv.org/abs/1811.03089}{{\ttfamily
  arXiv:1811.03089 [hep-ph]}}.

\bibitem{Gavela:2018paw}
M.~B. Gavela, M.~Ibe, P.~Quilez, and T.~T. Yanagida, ``{Automatic
  Peccei\textendash{}Quinn symmetry},''
  \href{http://dx.doi.org/10.1140/epjc/s10052-019-7046-3}{{\em Eur. Phys. J. C}
  {\bfseries 79} no.~6, (2019) 542},
  \href{http://arxiv.org/abs/1812.08174}{{\ttfamily arXiv:1812.08174
  [hep-ph]}}.

\bibitem{Vecchi:2021shj}
L.~Vecchi, ``{Axion quality straight from the GUT},''
  \href{http://dx.doi.org/10.1140/epjc/s10052-021-09745-x}{{\em Eur. Phys. J.
  C} {\bfseries 81} no.~10, (2021) 938},
  \href{http://arxiv.org/abs/2106.15224}{{\ttfamily arXiv:2106.15224
  [hep-ph]}}.

\bibitem{Contino:2021ayn}
R.~Contino, A.~Podo, and F.~Revello, ``{Chiral models of composite axions and
  accidental Peccei-Quinn symmetry},''
  \href{http://dx.doi.org/10.1007/JHEP04(2022)180}{{\em JHEP} {\bfseries 04}
  (2022) 180}, \href{http://arxiv.org/abs/2112.09635}{{\ttfamily
  arXiv:2112.09635 [hep-ph]}}.

\bibitem{Strassler:2006im}
M.~J. Strassler and K.~M. Zurek, ``{Echoes of a hidden valley at hadron
  colliders},'' \href{http://dx.doi.org/10.1016/j.physletb.2007.06.055}{{\em
  Phys. Lett. B} {\bfseries 651} (2007) 374--379},
  \href{http://arxiv.org/abs/hep-ph/0604261}{{\ttfamily arXiv:hep-ph/0604261}}.

\bibitem{Han:2007ae}
T.~Han, Z.~Si, K.~M. Zurek, and M.~J. Strassler, ``{Phenomenology of hidden
  valleys at hadron colliders},''
  \href{http://dx.doi.org/10.1088/1126-6708/2008/07/008}{{\em JHEP} {\bfseries
  07} (2008) 008}, \href{http://arxiv.org/abs/0712.2041}{{\ttfamily
  arXiv:0712.2041 [hep-ph]}}.

\bibitem{Arkani-Hamed:2008hhe}
N.~Arkani-Hamed, D.~P. Finkbeiner, T.~R. Slatyer, and N.~Weiner, ``{A Theory of
  Dark Matter},'' \href{http://dx.doi.org/10.1103/PhysRevD.79.015014}{{\em
  Phys. Rev. D} {\bfseries 79} (2009) 015014},
  \href{http://arxiv.org/abs/0810.0713}{{\ttfamily arXiv:0810.0713 [hep-ph]}}.

\bibitem{Kilic:2009mi}
C.~Kilic, T.~Okui, and R.~Sundrum, ``{Vectorlike Confinement at the LHC},''
  \href{http://dx.doi.org/10.1007/JHEP02(2010)018}{{\em JHEP} {\bfseries 02}
  (2010) 018}, \href{http://arxiv.org/abs/0906.0577}{{\ttfamily arXiv:0906.0577
  [hep-ph]}}.

\bibitem{Falkowski:2009yz}
A.~Falkowski, J.~Juknevich, and J.~Shelton, ``{Dark Matter Through the Neutrino
  Portal},'' \href{http://arxiv.org/abs/0908.1790}{{\ttfamily arXiv:0908.1790
  [hep-ph]}}.

\bibitem{Bai:2010qg}
Y.~Bai and R.~J. Hill, ``{Weakly Interacting Stable Pions},''
  \href{http://dx.doi.org/10.1103/PhysRevD.82.111701}{{\em Phys. Rev. D}
  {\bfseries 82} (2010) 111701},
  \href{http://arxiv.org/abs/1005.0008}{{\ttfamily arXiv:1005.0008 [hep-ph]}}.

\bibitem{Buckley:2012ky}
M.~R. Buckley and E.~T. Neil, ``{Thermal dark matter from a confining
  sector},'' \href{http://dx.doi.org/10.1103/PhysRevD.87.043510}{{\em Phys.
  Rev. D} {\bfseries 87} no.~4, (2013) 043510},
  \href{http://arxiv.org/abs/1209.6054}{{\ttfamily arXiv:1209.6054 [hep-ph]}}.

\bibitem{Boddy:2014yra}
K.~K. Boddy, J.~L. Feng, M.~Kaplinghat, and T.~M.~P. Tait, ``{Self-Interacting
  Dark Matter from a Non-Abelian Hidden Sector},''
  \href{http://dx.doi.org/10.1103/PhysRevD.89.115017}{{\em Phys. Rev. D}
  {\bfseries 89} no.~11, (2014) 115017},
  \href{http://arxiv.org/abs/1402.3629}{{\ttfamily arXiv:1402.3629 [hep-ph]}}.

\bibitem{Antipin:2015xia}
O.~Antipin, M.~Redi, A.~Strumia, and E.~Vigiani, ``{Accidental Composite Dark
  Matter},'' \href{http://dx.doi.org/10.1007/JHEP07(2015)039}{{\em JHEP}
  {\bfseries 07} (2015) 039}, \href{http://arxiv.org/abs/1503.08749}{{\ttfamily
  arXiv:1503.08749 [hep-ph]}}.

\bibitem{Harigaya:2016rwr}
K.~Harigaya and Y.~Nomura, ``{Light Chiral Dark Sector},''
  \href{http://dx.doi.org/10.1103/PhysRevD.94.035013}{{\em Phys. Rev. D}
  {\bfseries 94} no.~3, (2016) 035013},
  \href{http://arxiv.org/abs/1603.03430}{{\ttfamily arXiv:1603.03430
  [hep-ph]}}.

\bibitem{Co:2016akw}
R.~T. Co, K.~Harigaya, and Y.~Nomura, ``{Chiral Dark Sector},''
  \href{http://dx.doi.org/10.1103/PhysRevLett.118.101801}{{\em Phys. Rev.
  Lett.} {\bfseries 118} no.~10, (2017) 101801},
  \href{http://arxiv.org/abs/1610.03848}{{\ttfamily arXiv:1610.03848
  [hep-ph]}}.

\bibitem{Mitridate:2017oky}
A.~Mitridate, M.~Redi, J.~Smirnov, and A.~Strumia, ``{Dark Matter as a weakly
  coupled Dark Baryon},'' \href{http://dx.doi.org/10.1007/JHEP10(2017)210}{{\em
  JHEP} {\bfseries 10} (2017) 210},
  \href{http://arxiv.org/abs/1707.05380}{{\ttfamily arXiv:1707.05380
  [hep-ph]}}.

\bibitem{Contino:2018crt}
R.~Contino, A.~Mitridate, A.~Podo, and M.~Redi, ``{Gluequark Dark Matter},''
  \href{http://dx.doi.org/10.1007/JHEP02(2019)187}{{\em JHEP} {\bfseries 02}
  (2019) 187}, \href{http://arxiv.org/abs/1811.06975}{{\ttfamily
  arXiv:1811.06975 [hep-ph]}}.

\bibitem{Contino:2020god}
R.~Contino, A.~Podo, and F.~Revello, ``{Composite Dark Matter from
  Strongly-Interacting Chiral Dynamics},''
  \href{http://dx.doi.org/10.1007/JHEP02(2021)091}{{\em JHEP} {\bfseries 02}
  (2021) 091}, \href{http://arxiv.org/abs/2008.10607}{{\ttfamily
  arXiv:2008.10607 [hep-ph]}}.

\bibitem{Kribs:2016cew}
G.~D. Kribs and E.~T. Neil, ``{Review of strongly-coupled composite dark matter
  models and lattice simulations},''
  \href{http://dx.doi.org/10.1142/S0217751X16430041}{{\em Int. J. Mod. Phys. A}
  {\bfseries 31} no.~22, (2016) 1643004},
  \href{http://arxiv.org/abs/1604.04627}{{\ttfamily arXiv:1604.04627
  [hep-ph]}}.

\bibitem{Cline:2021itd}
J.~M. Cline, ``{Dark atoms and composite dark matter},''
  \href{http://dx.doi.org/10.21468/SciPostPhysLectNotes.52}{{\em SciPost Phys.
  Lect. Notes} {\bfseries 52} (2022) 1},
  \href{http://arxiv.org/abs/2108.10314}{{\ttfamily arXiv:2108.10314
  [hep-ph]}}.

\bibitem{Allanach:2021bfe}
B.~C. Allanach, B.~Gripaios, and J.~Tooby-Smith, ``{Semisimple extensions of
  the Standard Model gauge algebra},''
  \href{http://dx.doi.org/10.1103/PhysRevD.104.035035}{{\em Phys. Rev. D}
  {\bfseries 104} no.~3, (2021) 035035},
  \href{http://arxiv.org/abs/2104.14555}{{\ttfamily arXiv:2104.14555
  [hep-th]}}.

\bibitem{toappear}
R.~Contino, A.~Podo, and F.~Revello, ``{work in progress}''.

\bibitem{Witten:1982fp}
E.~Witten, ``{An SU(2) Anomaly},''
  \href{http://dx.doi.org/10.1016/0370-2693(82)90728-6}{{\em Phys. Lett. B}
  {\bfseries 117} (1982) 324--328}.

\bibitem{Witten:1985xe}
E.~Witten, ``{GLOBAL GRAVITATIONAL ANOMALIES},''
  \href{http://dx.doi.org/10.1007/BF01212448}{{\em Commun. Math. Phys.}
  {\bfseries 100} (1985) 197}.

\bibitem{Garcia-Etxebarria:2018ajm}
I.~n. Garc\'\i{}a-Etxebarria and M.~Montero, ``{Dai-Freed anomalies in particle
  physics},'' \href{http://dx.doi.org/10.1007/JHEP08(2019)003}{{\em JHEP}
  {\bfseries 08} (2019) 003}, \href{http://arxiv.org/abs/1808.00009}{{\ttfamily
  arXiv:1808.00009 [hep-th]}}.

\bibitem{Davighi:2019rcd}
J.~Davighi, B.~Gripaios, and N.~Lohitsiri, ``{Global anomalies in the Standard
  Model(s) and Beyond},'' \href{http://dx.doi.org/10.1007/JHEP07(2020)232}{{\em
  JHEP} {\bfseries 07} (2020) 232},
  \href{http://arxiv.org/abs/1910.11277}{{\ttfamily arXiv:1910.11277
  [hep-th]}}.

\bibitem{Wan:2020ynf}
Z.~Wan and J.~Wang, ``{Beyond Standard Models and Grand Unifications:
  Anomalies, Topological Terms, and Dynamical Constraints via Cobordisms},''
  \href{http://dx.doi.org/10.1007/JHEP07(2020)062}{{\em JHEP} {\bfseries 07}
  (2020) 062}, \href{http://arxiv.org/abs/1910.14668}{{\ttfamily
  arXiv:1910.14668 [hep-th]}}.

\bibitem{Mordell:1969}
L.~Mordell, {\em {Diophantine equations}}.
\newblock Pure and Applied Mathematics. Academic Press, 1969.

\bibitem{Dine:1981rt}
M.~Dine, W.~Fischler, and M.~Srednicki, ``{A Simple Solution to the Strong CP
  Problem with a Harmless Axion},''
  \href{http://dx.doi.org/10.1016/0370-2693(81)90590-6}{{\em Phys. Lett. B}
  {\bfseries 104} (1981) 199--202}.

\bibitem{Zhitnitsky:1980tq}
A.~R. Zhitnitsky, ``{On Possible Suppression of the Axion Hadron Interactions.
  (In Russian)},'' {\em Sov. J. Nucl. Phys.} {\bfseries 31} (1980) 260.

\bibitem{Kim:1979if}
J.~E. Kim, ``{Weak Interaction Singlet and Strong CP Invariance},''
  \href{http://dx.doi.org/10.1103/PhysRevLett.43.103}{{\em Phys. Rev. Lett.}
  {\bfseries 43} (1979) 103}.

\bibitem{Shifman:1979if}
M.~A. Shifman, A.~I. Vainshtein, and V.~I. Zakharov, ``{Can Confinement Ensure
  Natural CP Invariance of Strong Interactions?},''
  \href{http://dx.doi.org/10.1016/0550-3213(80)90209-6}{{\em Nucl. Phys. B}
  {\bfseries 166} (1980) 493--506}.

\bibitem{GrillidiCortona:2015jxo}
G.~Grilli~di Cortona, E.~Hardy, J.~Pardo~Vega, and G.~Villadoro, ``{The QCD
  axion, precisely},'' \href{http://dx.doi.org/10.1007/JHEP01(2016)034}{{\em
  JHEP} {\bfseries 01} (2016) 034},
  \href{http://arxiv.org/abs/1511.02867}{{\ttfamily arXiv:1511.02867
  [hep-ph]}}.

\end{thebibliography}\endgroup
}

\end{document}